\documentclass[a4paper,11pt]{article}
\pdfoutput=1 

\usepackage{jinstpub} 

\usepackage{lineno}
\usepackage[binary-units=true]{siunitx}

\usepackage{multirow}

\usepackage{textcomp}
\usepackage{textgreek}
\usepackage{xcolor}
\usepackage{graphicx}
\usepackage{mathtools}
\usepackage{subcaption}
\usepackage[LGRgreek]{mathastext}
\usepackage{xspace}
\usepackage{tikz}
\usetikzlibrary{arrows.meta} 

\usepackage{multirow}
\usepackage{adjustbox}
\usepackage{booktabs}
\usepackage[para]{threeparttable}
\usepackage{tabularx}
\usepackage{siunitx}
\newcolumntype{Y}{>{\centering\small\arraybackslash}X}

\newcolumntype{Z}{>{\hsize=.5\hsize}Y}
\newcolumntype{R}{>{\hsize=.3\hsize}Y}
\newcolumntype{G}{>{\hsize=.8\hsize}Y}
\newcolumntype{H}{>{\hsize=.5\hsize\raggedleft\small\arraybackslash}p{1.7cm}}

\newcommand{\ind}[0]{\hspace*{\customindent}}

\newcommand{\parag}{\parag\ind}

\newcommand{\tocite}[1]{\textbf{\textcolor{red}{[\ifthenelse{\equal{#1}{}}{?}{#1}]}}}

\newcommand{\percent}{$\%$}

\newcommand{\neqcm}{\ensuremath{\mathrm{n_{eq}}/\mathrm{cm}^{2}}\xspace}

\definecolor{ttHcol}{RGB}{67,118,201}
\definecolor{ttbbcol}{RGB}{235,230,10}
\definecolor{ttbcol}{RGB}{205,0,10}
\definecolor{tttbcol}{RGB}{255,153,1}
\definecolor{ttcccol}{RGB}{131,38,10}
\definecolor{ttlfcol}{RGB}{81,142,25}
\definecolor{Singletcol}{RGB}{0,204,204}
\definecolor{Vjetscol}{RGB}{104,140,140}
\definecolor{Dibosoncol}{RGB}{1,25,147}
\definecolor{ttVcol}{RGB}{255,102,101}
\definecolor{DarkGreen}{RGB}{0,122,0}

\title{\boldmath Leakage current of high-fluence neutron-irradiated 8”~silicon sensors for the CMS Endcap Calorimeter Upgrade}

\emailAdd{marta.adamina.sailer@cern.ch}

\abstract{
The HL-LHC will challenge the detectors with a nearly 10-fold increase in integrated luminosity compared to the previous LHC runs 
combined, thus the CMS detector will be upgraded to face the higher levels of radiation and the larger amounts of collision data to be collected. 
The High-Granularity Calorimeter will replace the current endcap calorimeters of the CMS detector. 
It will facilitate the use of particle-flow calorimetry with its unprecedented transverse and longitudinal readout/trigger segmentation, 
with more than 6M readout channels. The electromagnetic section as well as the high-radiation regions of the hadronic section of the HGCAL 
(fluences above $10^{14}~\neqcm$) will be equipped with silicon pad sensors, covering a total area of \SI{620}{\metre\squared}. Fluences up to $10^{16}~\neqcm$ and 
doses up to 1.5 MGy are expected. The sensors are processed on novel 8" p-type wafers with an active thickness of \SI{300}{\micro\metre}, 
\SI{200}{\micro\metre} and \SI{120}{\micro\metre}
and cut into hexagonal shapes for optimal use of the wafer area and tiling. Each sensor contains several hundred individually read out cells 
of two sizes (around \SI{0.6}{\centi\metre\squared} or \SI{1.2}{\centi\metre\squared}). To investigate the radiation-induced bulk damage, the sensors have been irradiated with 
neutrons at RINSC to fluences between $6.5 \times 10^{14}~\neqcm$ and $1.3 \times 10^{16}~\neqcm$. 
Electrical characterization results are presented for full sensors, as well as for partial sensors cut from multi-geometry wafers with internal dicing lines 
on the HV potential within the active sensor area. Leakage current behaviour is investigated for various sensor types and fluence levels, including its temperature dependence.
Finally, methods to limit the annealing time of the sensors during irradiation are investigated by analysing the impact of splitting high-fluence irradiations.
}


\collaboration{\includegraphics[height=17mm]{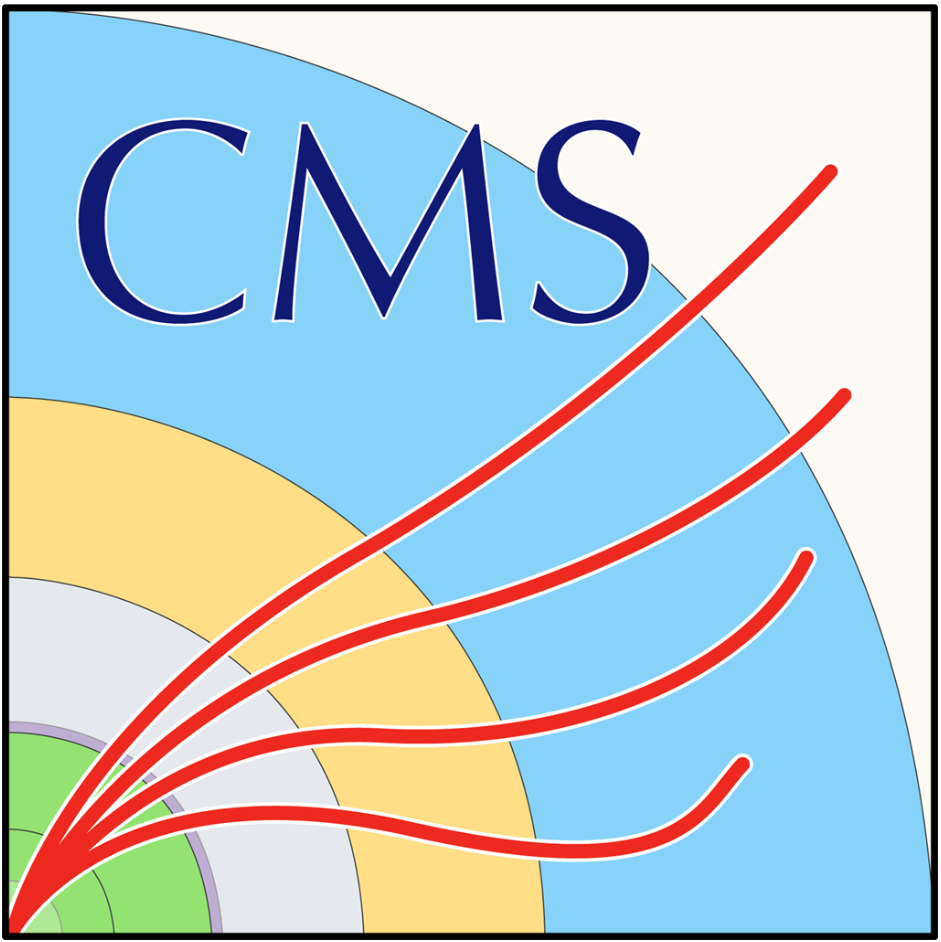}\\[6pt]
The CMS HGCAL collaboration}

\newcommand{\cmsorcid}[1]{\href{https://orcid.org/#1}{\raisebox{0.5ex}{\includegraphics[width=0.7em]{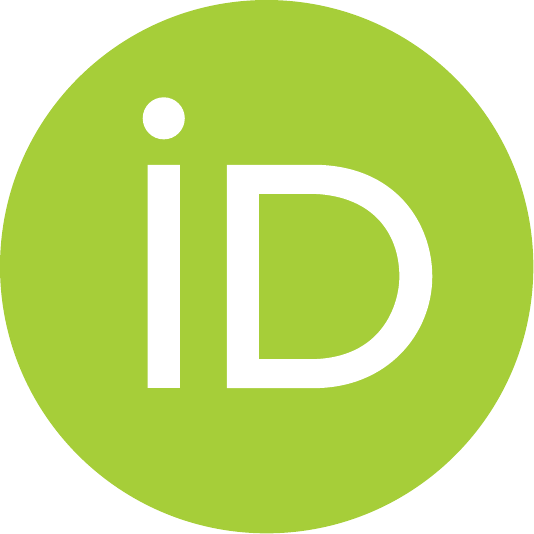}}}}

\author[60]{G.~Adamov~\cmsorcid{0009-0007-1832-8038}}
\author[23]{T.~Adams~\cmsorcid{0000-0001-8049-5143}}
\author[20]{S.~Afanasiev~\cmsorcid{0009-0006-8766-226X}}
\author[62]{C.~Agrawal~}
\author[30]{A.~Ahmad~\cmsorcid{0000-0002-4770-1897}}
\author[30]{H.~A.~Ahmed~\cmsorcid{0000-0002-3558-0928}}
\author[30]{S.~Akbar~}
\author[59]{N.~Akchurin~\cmsorcid{0000-0002-6127-4350}}
\author[66]{B.~Akgul~\cmsorcid{0000-0001-6621-8537}}
\author[10]{B.~Akgun~\cmsorcid{0000-0001-8888-3562}}
% ...
\affiliation[10]{Bo\u{g}azi\c{c}i University, \\Bebek, 34342 Istanbul, T\"{u}rkiye.}
% ...
\affiliation[20]{Authors affiliated with an international laboratory covered by a cooperation agreement with CERN}
% ...
\affiliation[23]{Florida State University, \\600 W. College Ave., Tallahassee 32306, FL, USA}
% ...
\affiliation[30]{National Centre for Physics, Quaid-I-Azam University, \\Islamabad-44000, Pakistan.}
% ...
\affiliation[59]{Texas Tech University, \\Department of Physics and Astronomy, Lubbock 79409, TX, USA}
\affiliation[60]{Georgian Technical University, \\Kostava str,  Tiiblisi, 77, 0160, Georgia.}
% ...
\affiliation[62]{Tata Inst. of Fundamental Research-B, \\Homi Bhabha Road, Mumbai 400005, India}
% ...
\affiliation[66]{Y{\i}ld{\i}z Technical University, \\Esenler, 34220 Istanbul, T\"{u}rkiye.}
% ...

\begin{document}
\renewcommand*{\thefootnote}{\fnsymbol{footnote}}
\footnotetext[1]{Corresponding author. Email: \text{marta.adamina.sailer@cern.ch}}
\maketitle
\renewcommand{\thefootnote}{\arabic{footnote}}
\setcounter{footnote}{0}
\flushbottom

\section{Introduction}

\label{sec:intro}

The Large Hadron Collider (LHC)~\cite{evans:2008} will enter its High-Luminosity phase (HL-LHC) in 2030.
It will challenge the detectors with a nearly 10-fold increase in integrated luminosity compared 
to the previous LHC runs combined~\cite{hl-lhc-tdr:2017}.
Therefore, the CMS~\cite{cms:2008} detector will be upgraded to handle the higher levels of radiation and 
the larger amounts of collision data to be collected. The CMS endcap calorimeter (CE), also 
known as High-Granularity Calorimeter (HGCAL), will replace 
the current endcap calorimeters of the CMS detector~\cite{hgcal-tdr:2018}. It will facilitate the use of particle-flow 
calorimetry with its unprecedented transverse and longitudinal readout/trigger segmentation, 
with more than 6M readout channels.
The electromagnetic section as well as the high-radiation regions of the hadronic section of the CE 
(fluences above $1 \times 10^{14}$ 1-MeV neutron-equivalents per square centimetre - $\neqcm$), will be equipped with silicon pad sensors, covering a total area of 620 m$^2$.
Integrated fluences up to $1 \times 10^{16}~\neqcm$ are expected at the end of its 10-year operation at the HL-LHC. 
The producer of the silicon sensors is Hamamatsu Photonics K.K.\footnote{\url{https://www.hamamatsu.com/eu/en.html}} The sensors are processed on novel 8'' p-type wafers with an active thickness of \SI{300}{\micro\metre}, \SI{200}{\micro\metre} and \SI{120}{\micro\metre} and cut into 
 hexagonal shapes for optimal use of the wafer area and tiling. Each sensor contains several hundred 
 individually read-out cells of two main sizes: around \SI{0.6}{\centi\metre\squared} for HD (High Density) sensors and \SI{1.2}{\centi\metre\squared} for LD (Low Density) sensors. In order to investigate the 
 radiation-induced bulk damage, the sensors have been irradiated with neutrons at RINSC (Rhode Island 
 Nuclear Science Centre, US) to target fluences between $6.5 \times 10^{14}~\neqcm$ and $1.3 \times 10^{16}~\neqcm$.

 This study includes the dataset analysed in ref.~\cite{Acar_2023} (version 1 campaign) together with the newly analysed version 2 campaign. 
 Compared with version 1, the version 2 sensors have an upgraded layout designed to improve high-voltage (HV) performance and to ensure reliable operation at higher fluence. 
 Accordingly, version 2 targeted a maximum fluence \SI{30}{\percent} above the $1 \times 10^{16}~\neqcm$ of version 1, in order to cover the uncertainty in the integrated fluence prediction for the final years of HL-LHC operation. 
 The version 2 campaign also includes additional sensor geometry variants.

  The results of this work have been used to optimize the layout of the
  CE silicon section and to deploy thicker sensors in higher radiation regions. Additionally, the irradiation facility RINSC is investigated for a better understanding and control
   of the irradiation processes and procedures
  with emphasis on high-fluence irradiations.

This paper is organized as follows: section~\ref{sec:Sensors} describes the version 2 prototypes of the 
CE silicon pad sensors, irradiated and measured in the campaign conducted after the publication of the previous study.
 Section~\ref{sec:Irradiation} focuses on the RINSC irradiation facility, describing the process improvements to reduce in-reactor annealing times. 
 Section~\ref{sec:LeakageCurrent} examines the leakage current distribution across the sensor area, with a particular focus on investigating potential effects related to the presence of internal guard rings and high-voltage lines in partial sensors.
Section~\ref{sec:Volume_normalized_current} deals with the relationship between leakage current and voltage, presenting key observations, including exponential leakage current behaviour under specific conditions.
The leakage current-related damage rate coefficient ($\alpha$) for various sensor types and fluence levels is presented in section~\ref{sec:Current_related_damage_rate}, along with an evaluation of the fluence estimation methods applied during irradiation. Section~\ref{sec:Temperature} presents the 
temperature dependence of the leakage current to extract the silicon activation energy and identify the primary source of the leakage current. The conclusions of this study are summarized in section \ref{sec:Summary}.
\section{Silicon sensors for the CMS endcap calorimeter upgrade}

\label{sec:Sensors}

The previous study~\cite{Acar_2023} described the fabrication and design parameters of the silicon sensors for the CE project, focusing on full sensors from the version 1 prototype campaign.
In contrast, the present study analysed version 2 prototypes introduced in 2021. 
An overview of the sensors examined in both campaigns is provided in table~\ref{tab:summary}.
The table summarizes the sensor characteristics, including layout, thickness, and the delivered fluence (see section~\ref{subsec:Fluence_assessment}). 
The term "epi" refers to the epitaxial growth process, while "FZ" stands for float-zone process.
The table also lists the material used for the irradiation holder (puck), 
and provides in-reactor annealing estimates derived from resistance temperature detector (RTD) measurements (see section~\ref{subsec:Annealing_assessment}).
\begin{table}[htbp]
    \centering
    \caption{Overview of the irradiation rounds performed at RINSC.}
    \label{tab:summary}
    \begin{threeparttable}
    \begin{tabularx}{\textwidth}{RZYYZYYGHHH}
        \toprule{}
        \footnotesize Ver\-sion & \footnotesize Round (Part) & \footnotesize Thick\-ness [$\upmu{}$m] &  Process & Lay\-out & Target fluence [$~\neqcm$] & Delivered fluence [$\neqcm$]\tnote{A} & Puck material & \multicolumn{3}{p{3cm}}{\centering In-reactor annealing [min at \SI{60}{\celsius}]\tnote{B}} \\
        \cmidrule(lr){9-11}
        &&&&&&&& Front\tnote{C} & Back & Average\tnote{D}\\
        \midrule
        1 & 1 & 300 & FZ & full & 6.5E+14 & \num{5.03E+14} & wood & 9.9 & -\tnote{E} & - \\
        1 & 2 & 200 & FZ & full & 2.5E+15 & \num{1.78E+15} & wood & 8.6 & -\tnote{E} & - \\
        1 & 3 & 120 & epi & full & 1.0E+16 & \num{8.07E+15} & wood & 2199.3 & -\tnote{E} &  \\ 
        1 & 4 & 200 & FZ & full & 2.5E+15 & \num{1.75E+15} & Acrylic & -\tnote{F} & -\tnote{E} & - \\
        1 & 5 & 200 & FZ & full & 2.5E+15 & \num{1.76E+15} & Acrylic & -\tnote{F} & -\tnote{E} & - \\
        1 & 6 & 120 & epi & full & 1.0E+16 & \num{7.84E+15} & Acrylic & -\tnote{G} & -\tnote{E} & - \\
        1 & 7 & 120 & epi & full & 2.5E+15 & \num{1.81E+15} & Acrylic & 1.3\tnote{H} & 1.1\tnote{H} & 1.2 \\
        1 & 8 & 120 & epi & full & 5.0E+15 & \num{3.48E+15} & Acrylic & 8.4 & 7.9 & 8.1 \\
        1 & 9 & 300 & FZ & full & 1.5E+15 & \num{1.08E+15} & Acrylic & 7.5 & 6.5 & 7.0 \\
        1 & 10 & 300 & FZ & full & 1.0E+15 & \num{7.20E+15} & Acrylic & 10.1 & 9.9 & 10.0 \\
        1 & 11 & 200 & FZ & full & 2.5E+15 & \num{1.91E+15} & Acrylic & 41.6 & 9.5 & 17.8\\
        \cline{1-11}
        2 & 1 & 300 & FZ & full & 1.5E+15 & \num{1.50E+15}  & Acrylic & 143.7 & 6.2 & 6.7\\
        2 & 2 & 300 & FZ & full & 2.0E+15 & \num{2.06E+15} & Acrylic & 837.5 & 3.6 & 29.7 \\
        2 & 3 & 200 & FZ & full & 4.0E+15 & \num{3.70E+15} & PEEK & 503.4 & 3.6 & 7.1 \\
        2 & 4 & 120 & epi & full & 1.0E+16 & \num{9.40E+15} & PEEK & >10000 & >10000 & >10000\tnote{I} \\
        2 & 5 & 200 & FZ & full & 4.0E+15 & \num{4.15E+15} & PEEK & 48.7 & 7.7 & 14.0 \\
        2 & 6 & 200 & FZ & full & 5.5E+15 & \num{5.22E+15} & PEEK & 1028.7 & 16.5 & 131.9 \\
        2 & 7 & 300 & FZ & full & 2.0E+15 & \num{2.06E+15} & Acrylic & 8.3 & 8.1 & 7.6 \\
        2 & 8 & 300 & FZ & full & 1.5E+15 & \num{1.55E+15} & Acrylic & 9.9 & 6.7 & 6.7 \\
        \cline{2-11}
        \multirow{2}{*}{2} & 9~(I)  & \multirow{2}{*}{120}& \multirow{2}{*}{epi} & \multirow{2}{*}{full} & \multirow{2}{*}{1.0E+16} & \num{5.38E+15} & \multirow{2}{*}{PEEK} & 126.2 & 8.3 & 18.7 \\ 
            & 9~(II) & & & & & \num{5.04E+15} & & 126.2\tnote{J} & 8.3\tnote{J} & 18.7\tnote{J} \\
            \cline{2-10}
            \multirow{2}{*}{2} & 10~(I) & \multirow{2}{*}{120} & \multirow{2}{*}{epi} & \multirow{2}{*}{full} & \multirow{2}{*}{1.3E+16} & \num{6.48E+15} & \multirow{2}{*}{Alu.\tnote{M}} & 88.7 & 36.3 & 54.8\\  
            & 10~(II) &  &  & & & \num{6.58E+15} &  & >10000\tnote{K} & - & - \\  
            \cline{2-10}
            2 & 11 & 200 & FZ & full & 5.5E+15 & \num{4.93E+15} & Alu.\tnote{M} & 124.1 & 70.9 & 80.2 \\  
        \cline{2-11}
        \multirow{2}{*}{2} & 12~(I) & \multirow{2}{*}{120} & \multirow{2}{*}{epi} & \multirow{2}{*}{full} & \multirow{2}{*}{1.3E+16} & \num{6.87E+15} & \multirow{2}{*}{Alu.\tnote{M}} & 64.3& 48.7 & 53.4\\  
        & 12~(II) & && && \num{6.62E+15} & & 121.4 & 36.8 & 62.6\\  
        \cline{2-11}
        2 & 13 & 200 & FZ & partial & 5.3E+15 & \num{5.25E+15} & Alu.\tnote{M} & 271.3 & 39.1 & 81.7\\  
        \cline{2-11}
        \multirow{2}{*}{2} & 14~(I) & \multirow{2}{*}{120} & \multirow{2}{*}{epi}& \multirow{2}{*}{partial}  & \multirow{2}{*}{1.0E+16} & \num{4.68E+15} & \multirow{2}{*}{Alu.\tnote{M}} & 52.4 & 24.5 & 30.2\\  
        & 14~(II) & & && & \num{4.82E+15} & & 34.9  & 24.5\tnote{L} & 30.2\tnote{L} \\  
    \bottomrule{}
    \end{tabularx}
    \vspace{-0.5cm}
    \begin{tablenotes}\footnotesize
    \item[A] Fluence evaluated from the irradiation time.
    \item[B] The in-reactor annealing is an estimate of total equivalent time at \SI{60}{\celsius}, RTD info.
    \item[C] The front side estimation is an average of the front RTD measurements if available. 
    \item[D] The average is assumed to be in the puck middle near sensor location.
    \item[E] No RTDs placed in the back.
    \item[F] Recording stopped before irradiation end, before the high temperature regime.
    \item[G] Puck underwent material degradation, no temperature recording is available. 
    \item[H] Recording stopped after irradiation end at \SI{13}{\celsius} (111 minutes).
    \item[I] Annealed into the reverse annealing regime, as defined in~\cite{moll:SiDamages,MOLL199987}.
    \item[J] Estimation from version 2, round 9 (I).
    \item[K] Annealed into the reverse annealing regime: 85-90$~\%$ of ice was available at the start of the irradiation round.
    \item[L] Estimation from version 2, round 14 (I).
    \item[M] Alu. = Aluminum
    \end{tablenotes}
    \end{threeparttable}
\end{table}

Between version 1 and version 2, several design updates were implemented to improve HV stability. 
These modifications included changes in the spacing between structures separating individual cells, as well as adjustments to critical HV structures at the sensor edges. The bulk material specifications remained unchanged between the two versions.

\begin{figure}
	\captionsetup[subfigure]{aboveskip=-1pt,belowskip=-1pt}
	\centering
	\begin{subfigure}[c]{0.49\textwidth}
		\centering
		\begin{tikzpicture}
            \node[anchor=south west,inner sep=0] (image1) at (0,0) {\includegraphics[width=0.85\textwidth]{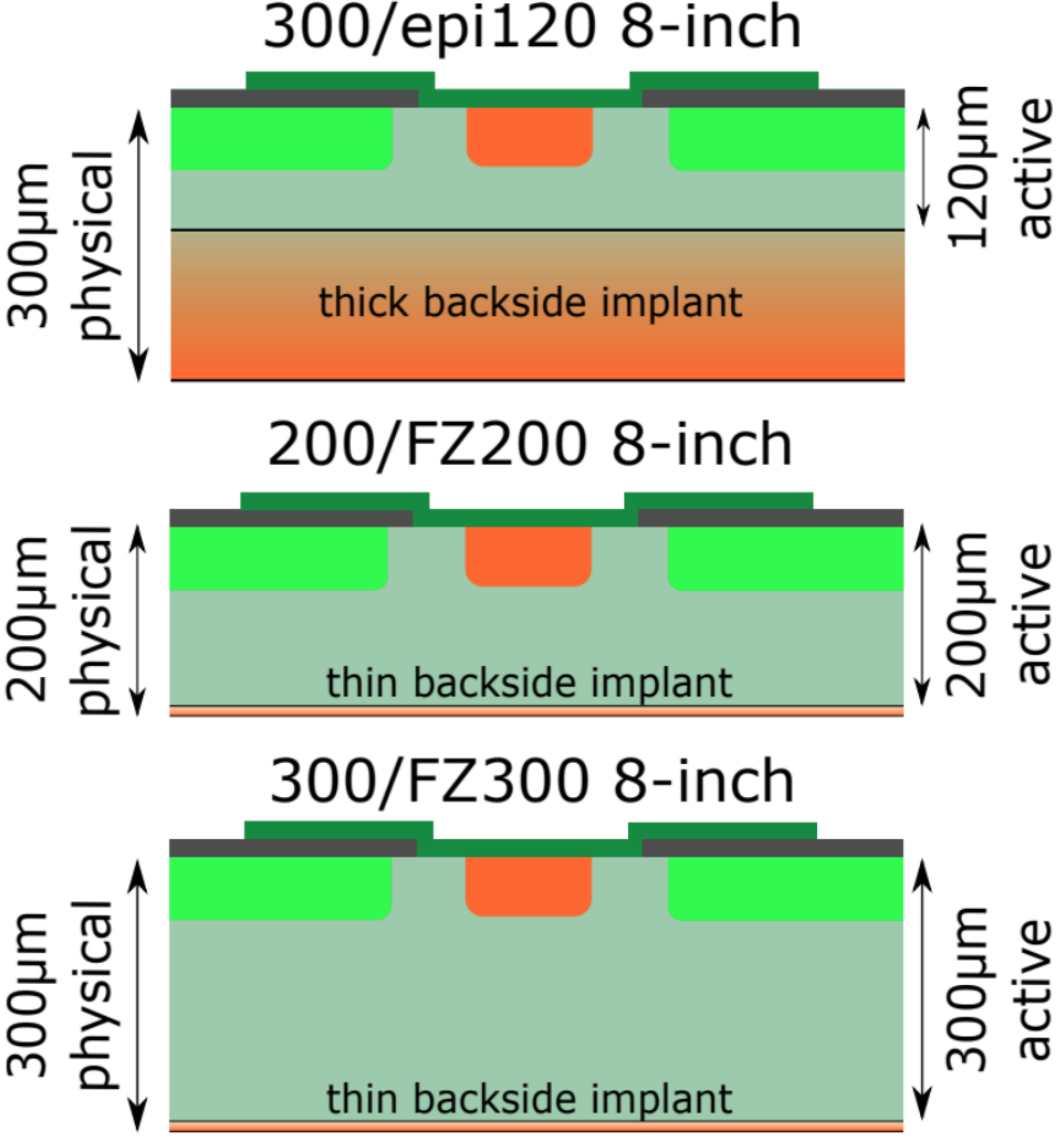}};
            \begin{scope}[x={(image1.south east)},y={(image1.north west)}]
            \end{scope}
        \end{tikzpicture}
		\subcaption{
		}
        \label{plot:Sensor_thickness}
	\end{subfigure}
	\hfill
	\begin{subfigure}[c]{0.5\textwidth}
		\centering
		\vspace{0.86cm}
		\begin{tikzpicture}
            \node[anchor=south west,inner sep=0] (image1) at (0,0) {\includegraphics[width=1.2\textwidth,trim=10 0 10 5, clip]{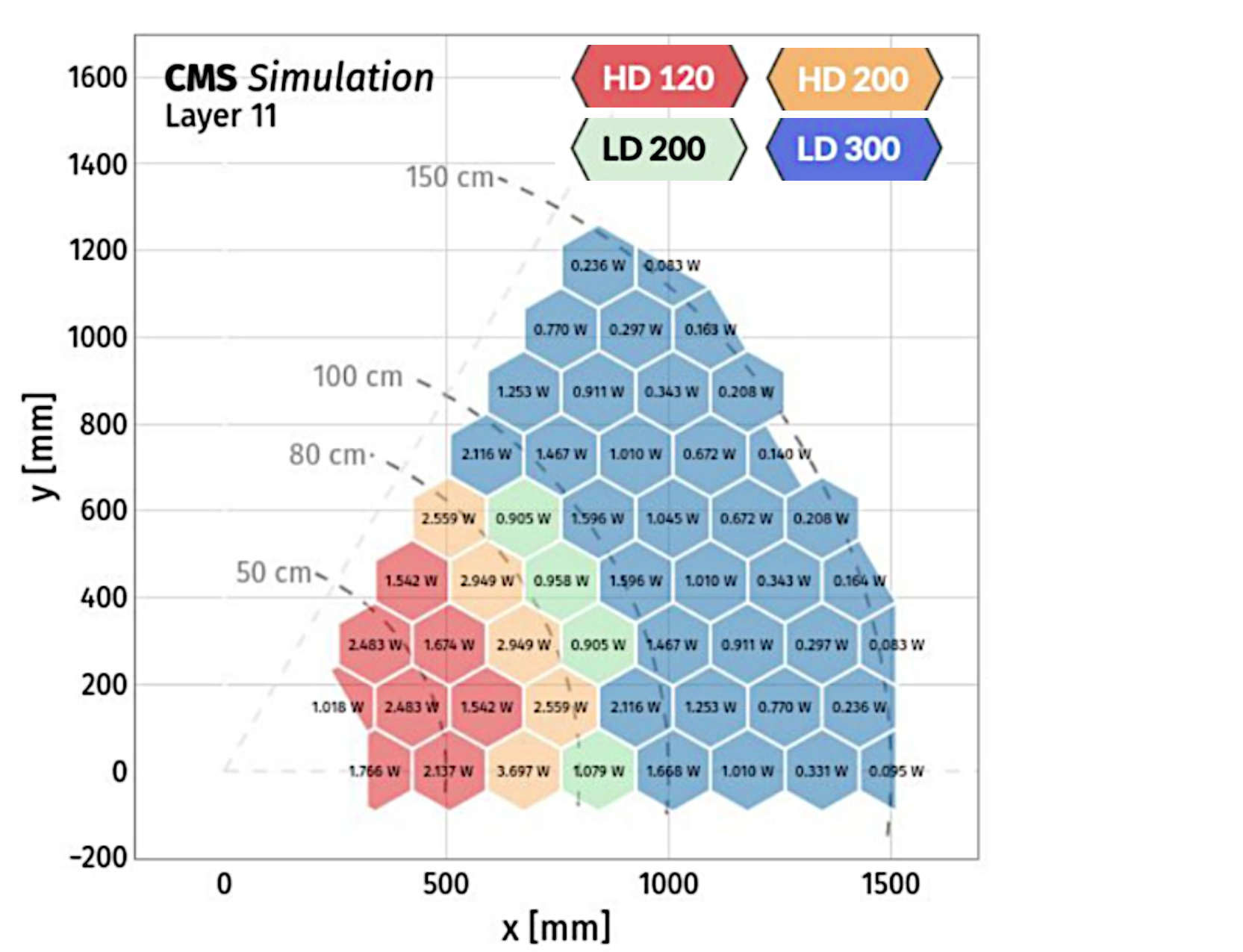}};
            \begin{scope}[x={(image1.south east)},y={(image1.north west)}]
            \end{scope}
        \end{tikzpicture}
		\subcaption{
		}
        \label{plot:Layout_Optimization}
	\end{subfigure}
	\newline
	\begin{subfigure}[c]{0.49\textwidth}
		\centering
		\begin{tikzpicture}
            \node[anchor=south west,inner sep=0] (image1) at (0,0) {\includegraphics[width=0.999\textwidth]{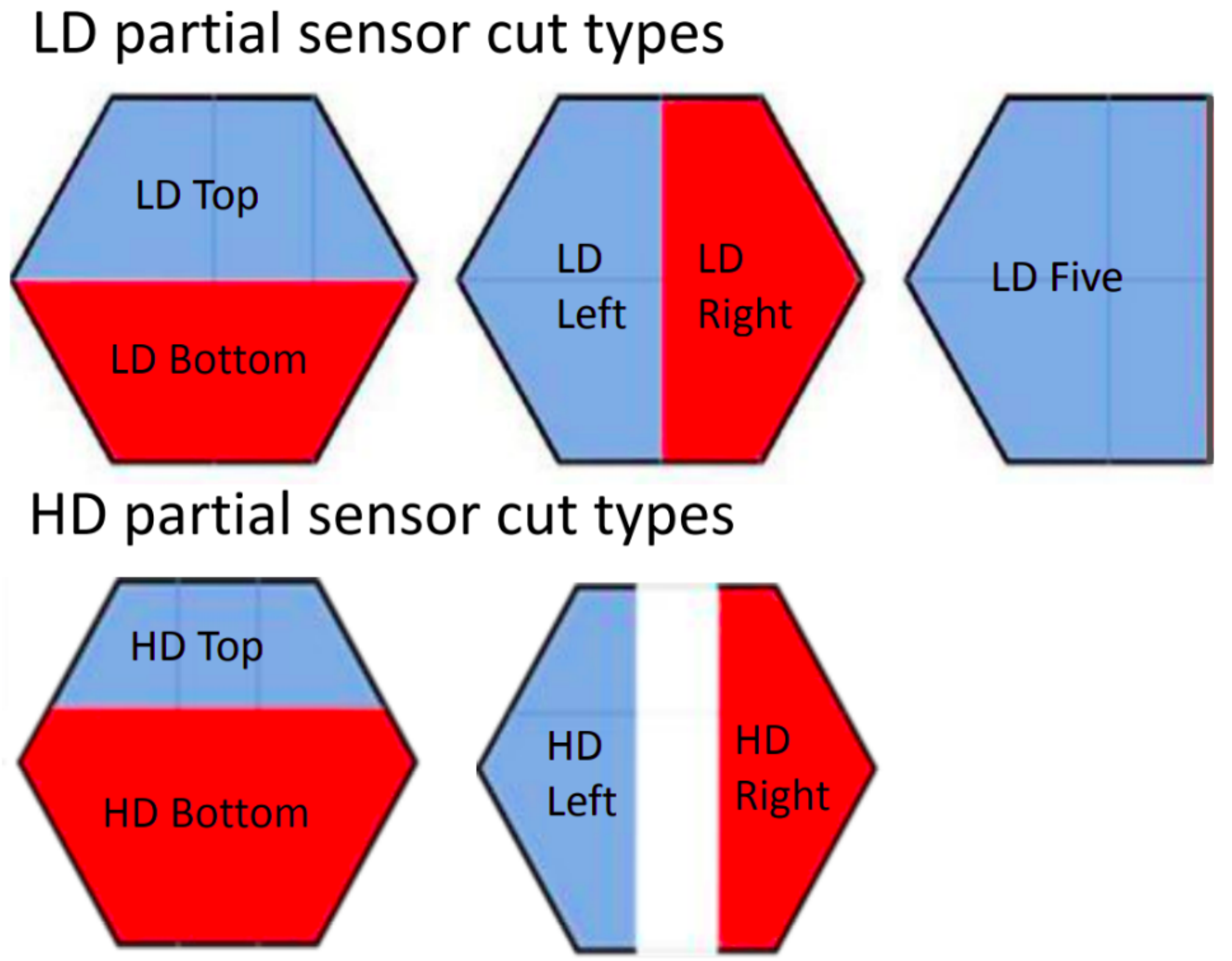}};
            \begin{scope}[x={(image1.south east)},y={(image1.north west)}]
            \end{scope}
        \end{tikzpicture}
		\subcaption{
		}
        \label{plot:Cut_types}
	\end{subfigure}
	\begin{subfigure}[c]{0.50\textwidth}
		\centering
		\includegraphics[width=0.70\textwidth]{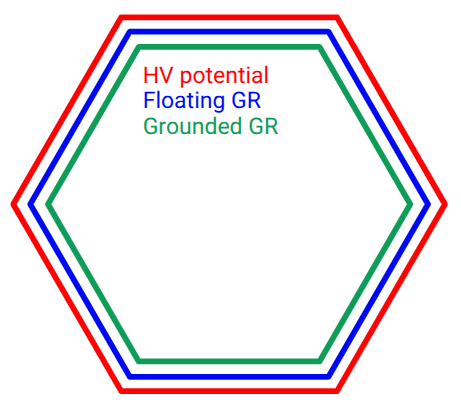}
		\subcaption{
		}		
        \label{plot:HV_full}
	\end{subfigure}
	\newline
	\begin{subfigure}[c]{0.49\textwidth}
		\centering
		\begin{tikzpicture}
			\node[anchor=south west,inner sep=0, xshift=-1cm, yshift=-9cm] (image1) at (0,0) {\includegraphics[width=1\textwidth]{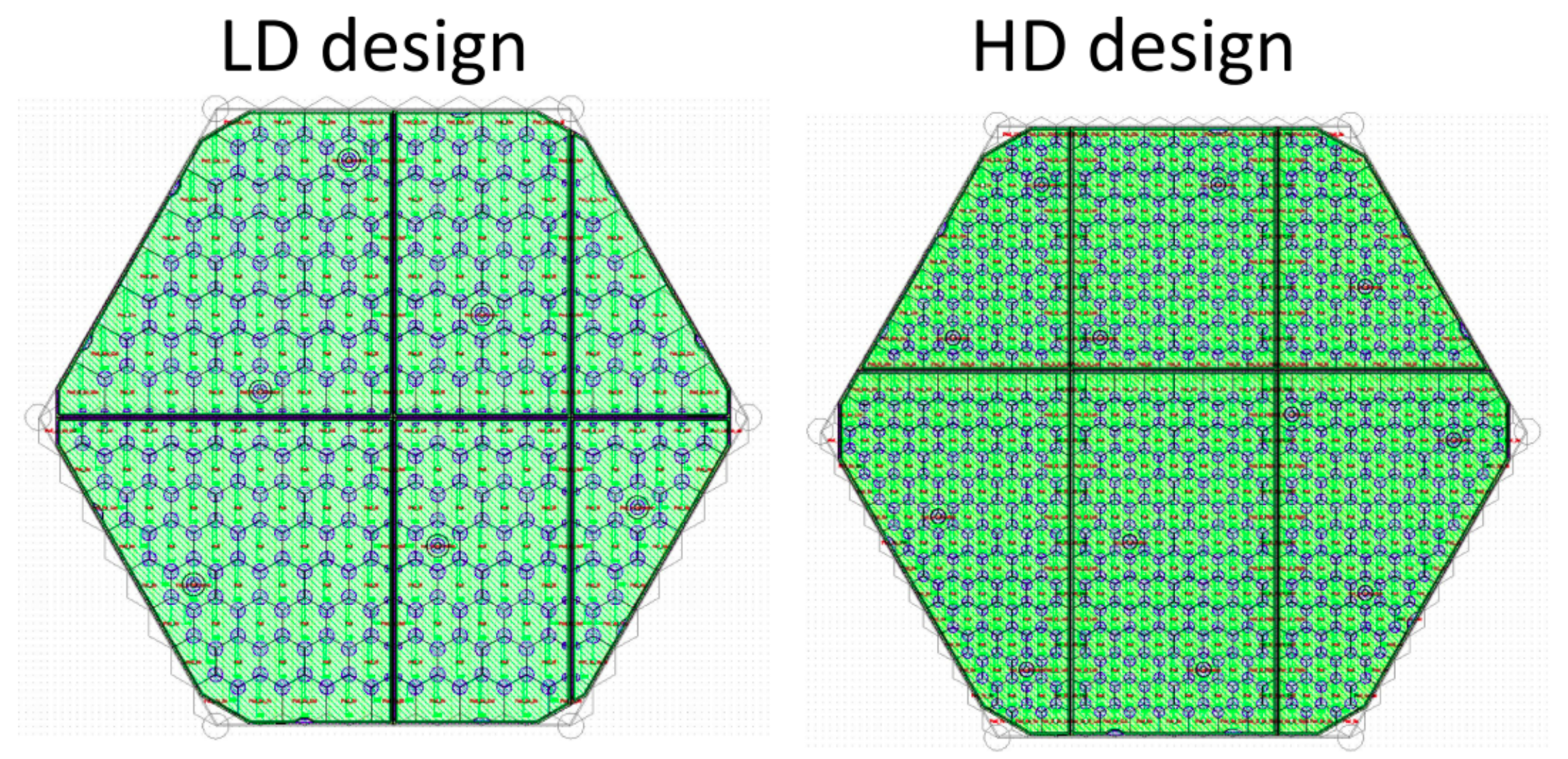}};
            \begin{scope}[x={(image1.south east)},y={(image1.north west)}]
				\node[anchor=south east, xshift=1.5cm, yshift=2cm, text=violet!90] at (0.5,0.5) {Internal dicing lines};
				\draw[violet!90, ultra thick, -{Latex}] (0.5,0.13) -- (0.15,1.1);
				\draw[violet!90, ultra thick, -{Latex}] (0.5,0.13) -- (0.28,0.85);
				\draw[violet!90, ultra thick, -{Latex}] (0.5,0.13) -- (0.38,0.65);
				\draw[violet!90, ultra thick, -{Latex}] (0.5,0.13) -- (0.74,0.08);
				\draw[violet!90, ultra thick, -{Latex}] (0.5,0.13) -- (0.67,0.35);
				\draw[violet!90, ultra thick, -{Latex}] (0.5,0.13) -- (0.81,0.23);
            \end{scope}

			\draw[black, thick, {Latex}-{Latex}] (-0.8, -8.85) -- (-0.8, -5.9);

			\node[anchor=south, rotate=90] at (-0.8, -7.3) {\SI{165}{\milli\metre}};

        \end{tikzpicture}
		\subcaption{
		}
        \label{plot:Partial_design}
	\end{subfigure}
	\hfill
	\begin{subfigure}[c]{0.50\textwidth}
		\centering
		\hspace*{-1.6cm}
		\includegraphics[width=0.55\textwidth]{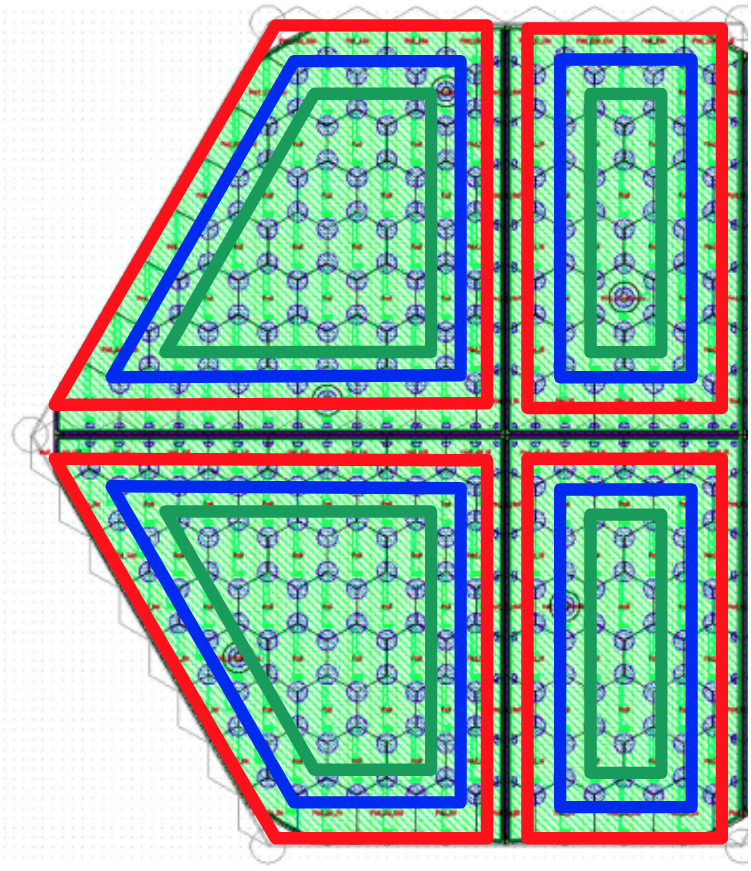}
		\subcaption{
		}		
        \label{plot:HV_partial}
	\end{subfigure}
	\hfill
	
	\caption{
		a) Sensor cross-sections for different manufacturing processes~\cite{Paulitsch:2020hci}. b) CMS simulation showing an optimized lateral layout for CE layer 11. c) Overview of LD and HD partial sensor cut types. d) Schematic (not to scale) 
		of the HV protection structures for a full sensor. e) Layouts of LD and HD 
        partial sensor design. f) Schematic (not to scale) of the HV protection structures for a partial sensor (LD Five), which are also present within the sensor's interior due to the internal dicing lines.
		}
	\label{plot:Partials}
\end{figure}

Three sensor thicknesses were deployed: \SI{300}{\micro\metre}, \SI{200}{\micro\metre} and \SI{120}{\micro\metre}, as illustrated in figure~\ref{plot:Sensor_thickness}. 
The \SI{300}{\micro\metre} and \SI{200}{\micro\metre} sensors were fabricated using the FZ process, while the \SI{120}{\micro\metre} sensors were produced using the epitaxial process 
on top of a \SI{180}{\micro\metre} thick handling wafer.
Radiation-induced defects in the bulk material generate leakage current, which increases with fluence.
Thinner sensors, having a smaller bulk volume, generate less leakage current and exhibit greater radiation 
hardness compared to thicker sensors~\cite{Moll:300958}.
As a result, thinner sensors are deployed in regions with higher radiation exposure, as illustrated in figure~\ref{plot:Layout_Optimization}.

In addition to the full sensors, so-called partial sensors were designed and produced from multi-geometry wafers (MGW) 
to cover border regions of the CE detector, as depicted in figure~\ref{plot:Layout_Optimization}. 
To maximize detector coverage, partial sensors were designed in various cut types (figure~\ref{plot:Cut_types}) and are available in both LD and HD granularities.
For this study, HD Bottom and LD Five cut types were selected, due to their ability to remain securely in place within the puck which was originally designed for full-sensor irradiations.

In the context of this study, the main difference between full and partial sensors lies in their design for manufacturing. 
A full sensor is fabricated from an entire silicon wafer, so the dicing lines, where the wafer is cut to define the sensor edges, 
are located only at the sensor periphery.
To protect the inner sensor cells from the high voltage at the edges, dedicated HV protection structures are implemented exclusively at the dicing lines, 
as shown schematically in figure~\ref{plot:HV_full}.

The HV ring is located at the sensor edge and is biased to the applied high voltage. 
Adjacent to the HV ring are two guard rings (GR), which provide the HV protection. 
The first guard ring (blue) is floating, meaning that it is not connected to any fixed potential. 
The innermost guard ring (green) is connected to ground potential and shields the inner sensor cells from the high voltage at the sensor edge.

In contrast, partial sensors are manufactured by cutting multiple sensor types from a single wafer (figure~\ref{plot:Cut_types}). 
As a result, internal dicing lines are introduced within the sensor layout, as shown in figure~\ref{plot:Partial_design}. 
To protect the sensor interior from the high voltage present within the sensor area, additional HV protection structures are required.

\begin{figure}
	\captionsetup[subfigure]{aboveskip=-1pt,belowskip=-1pt}
	\centering
    
	\begin{subfigure}[c]{0.49\textwidth}
		\centering
		
		\begin{tikzpicture}
            \node[anchor=south west,inner sep=0] (image1) at (0,0) {\includegraphics[width=0.7\textwidth]{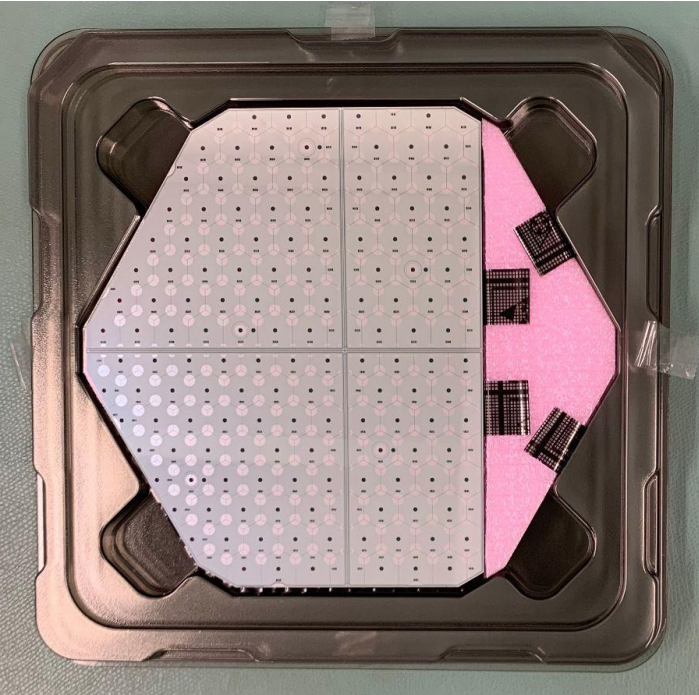}};

			\def\imgW{5.0}  
			\def\imgH{5.0}  
	
			\draw[white, thick, {Latex}-{Latex}] (1.55,\imgH-0.75) -- (\imgW-1.4,0.9);
	

			\node[white, rotate=-58] at (2.75,2.7) {\SI{18}{\centi\metre}};

        \end{tikzpicture}
		\vspace{0.55cm}
		\subcaption{
		}
        \label{plot:LD_Five}
	\end{subfigure}
	\hfill
	\begin{subfigure}[c]{0.49\textwidth}
		\centering
		\begin{tikzpicture}
			\node[anchor=south west,inner sep=0] (image1) at (-7,0) 
				{\includegraphics[width=0.7\textwidth]{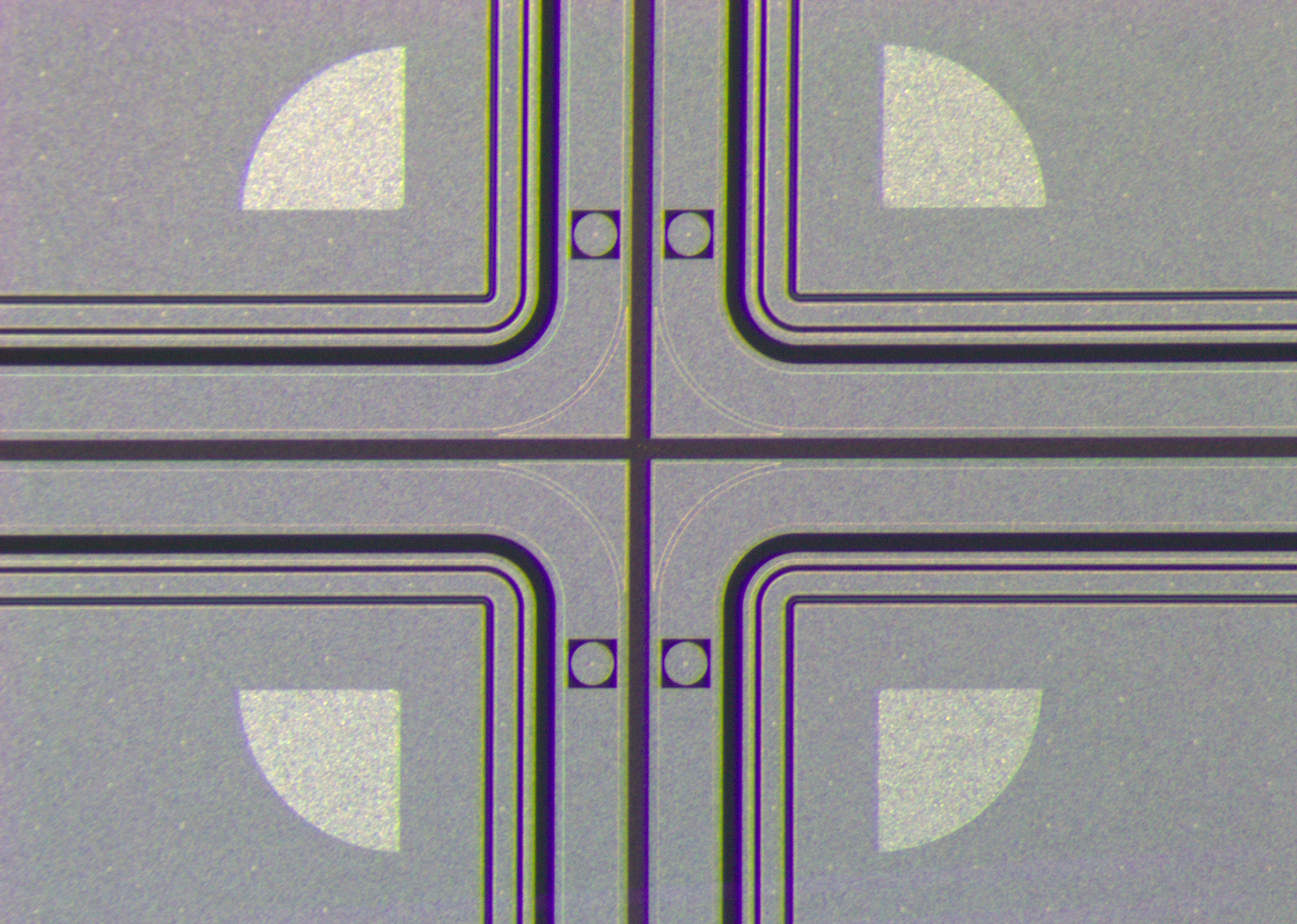}};
			
			\node[anchor=south, yshift=-0.6cm, text=black, rotate=90] 
				at (-7.15,2.5) {\SI{1}{\milli\metre}};
		
			\draw[black, thick, {Latex}-{Latex}] (-7.1, 1.5) -- (-7.1, 2.3);


			\draw[violet!90, thick, -{Latex}] (-6, 3.8) -- (-4.4, 3.0);
			\draw[violet!90, thick, -{Latex}] (-6, 3.8) -- (-6.5, 1.9);

			\node[align=center, text=violet!90] at (-6.3, 4.2) {Internal\\dicing lines};

			\draw[black, thick, -{Latex}] (-3, 3.8) -- (-3.15, 3.1);

			\node[align=center, text=black] at (-3.3, 4.2) {Cell\\contact pad};

			\draw[red, thick, -{Latex}] (-6, -0.1) -- (-4.2, 2.4);

			\node[align=center, text=red] at (-6.3, -0.5) {Internal\\HV  ring};

			\draw[blue, thick, -{Latex}] (-4.5, -0.1) -- (-3.9, 2.38);

			\node[align=center, text=blue] at (-4.1, -0.5) {Floating\\GR};

			\draw[DarkGreen, thick, -{Latex}] (-2, -0.1) -- (-3.5, 2.45);

			\node[align=center, text=DarkGreen] at (-2., -0.5) {Grounded\\GR};
		\end{tikzpicture}

		\subcaption{
		}
        \label{plot:Internal_HV}
	\end{subfigure}
	\hfill
	\label{plot:HV_lines}
	\caption{
		a) Photograph of an LD Five partial sensor placed in the logistics tray. 
		b) Microscope image showing the internal dicing lines, HV rings, and internal HV protection structures (including the floating GR and grounded GR), on a partial sensor.
		}
\end{figure}

A schematic representation of the HV protection structures in a partial sensor of type LD Five is shown in figure~\ref{plot:HV_partial}.
It illustrates how both HV rings and internal GR structures (floating and grounded) are integrated within the sensor interior.
A photograph of a sensor of cut type LD Five is provided in figure~\ref{plot:LD_Five}, while figure~\ref{plot:Internal_HV} shows a detailed view 
of the intersection of HV rings and internal HV protection structures in the inner sensor area.

\begin{figure}
    \captionsetup[subfigure]{aboveskip=-1pt,belowskip=-1pt}
    \centering

    \begin{subfigure}[c]{0.45\textwidth}
        \centering
        \begin{tikzpicture}
            \node[anchor=south west, inner sep=0] (image) at (0,0)
                {\includegraphics[width=0.7\textwidth, angle=270]{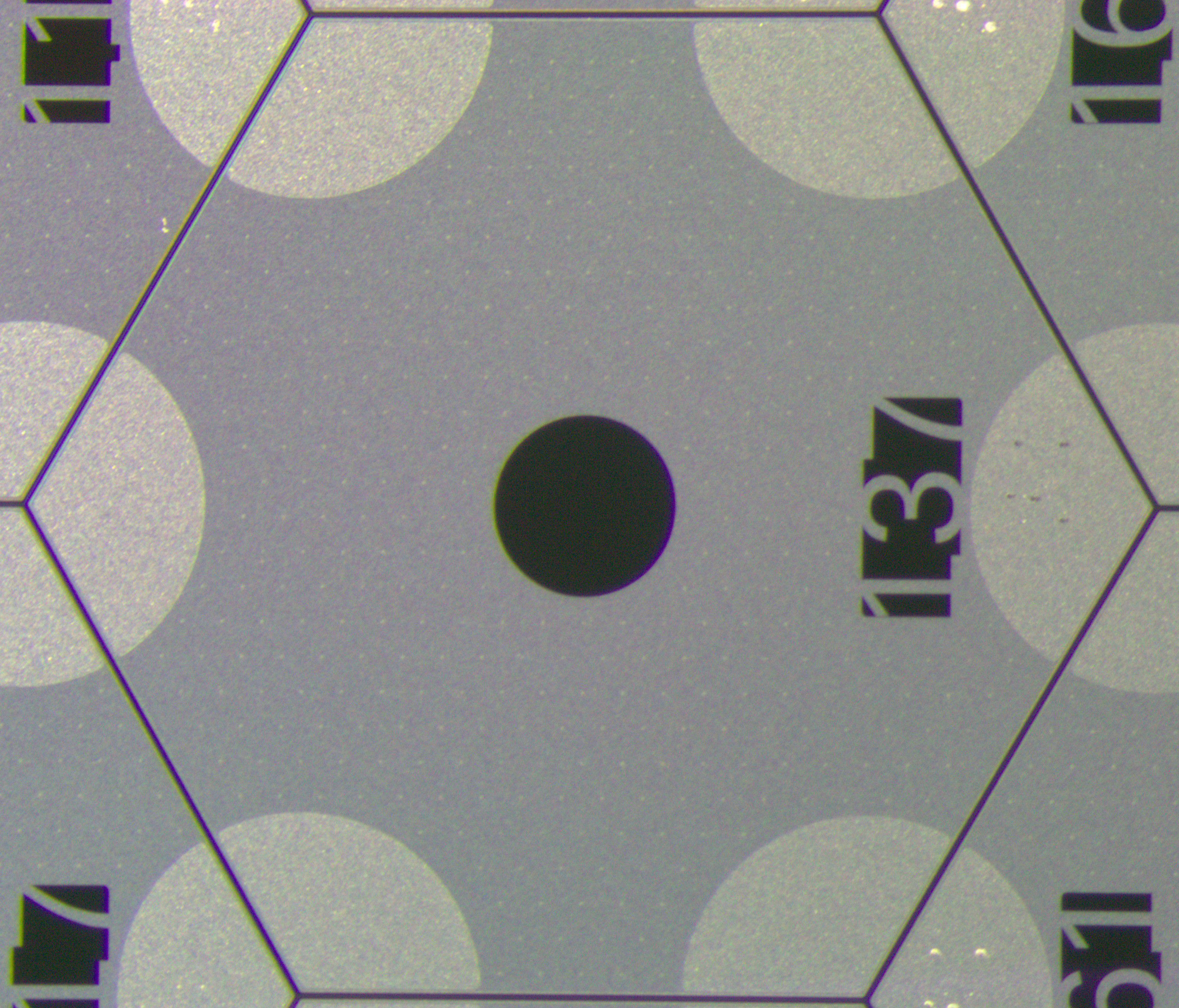}};
			\draw[white, thick] (0.06,0.5) -- ++(0.75,0);
            \node[text=white, anchor=north] at (0.56,0.5) {\footnotesize \SI{1.5}{\milli\metre}};
			\draw[green, thick] (2.1,0.53) circle [radius=0.3];
        \end{tikzpicture}
        \subcaption{}
        \label{plot:full_cell}
    \end{subfigure}
    \hfill
    \begin{subfigure}[c]{0.45\textwidth}
        \centering
		\begin{tikzpicture}
            \node[anchor=south west, inner sep=0] (image) at (0,0)
                {\includegraphics[width=0.7\textwidth, angle=270]{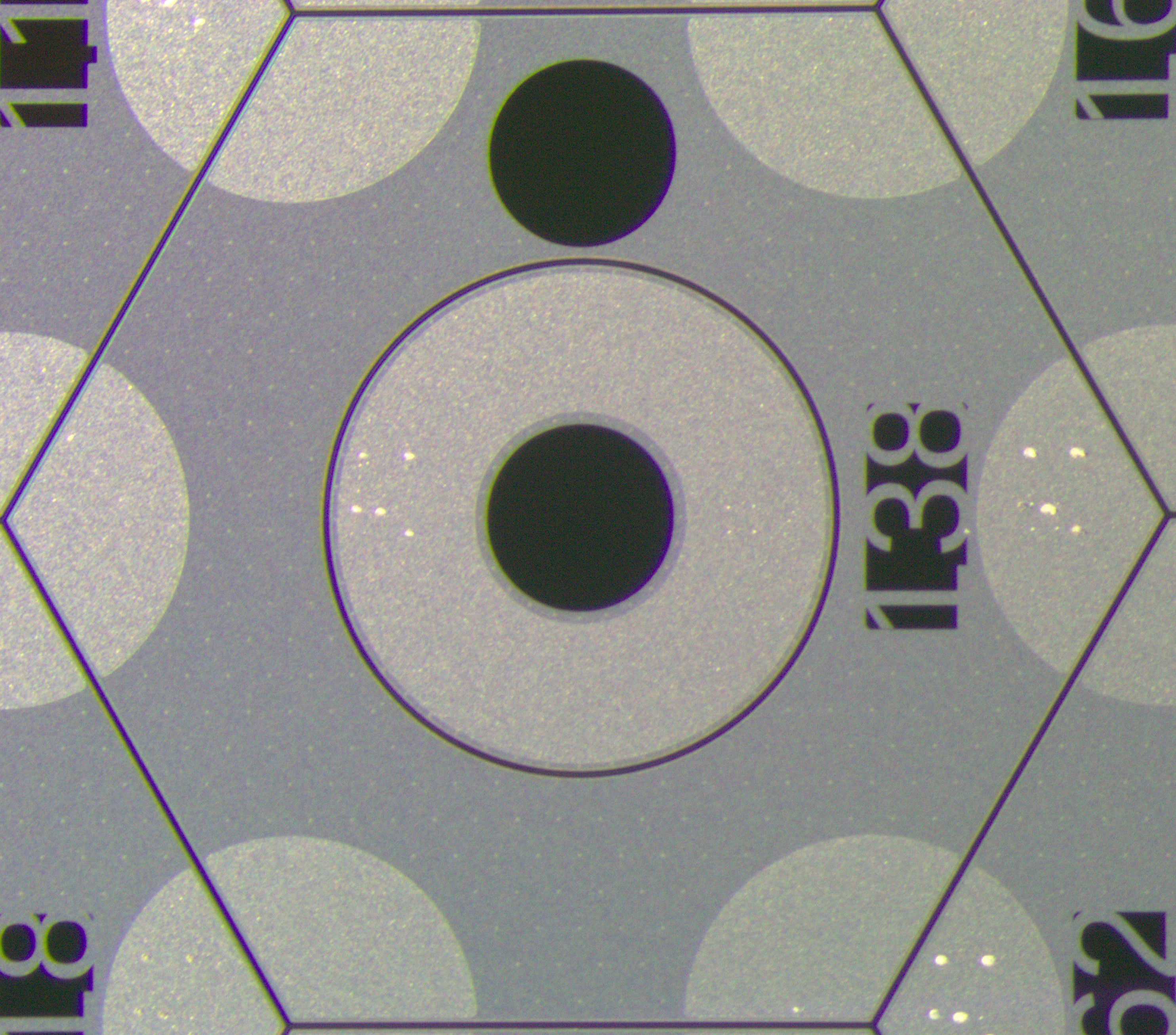}};
			\draw[white, thick] (0.06,0.5) -- ++(0.75,0);
			\node[text=white, anchor=north] at (0.56,0.5) {\footnotesize \SI{1.5}{\milli\metre}};
			\draw[green, thick] (2.18,3.18) circle [radius=0.3];
			\draw[green, thick] (2.18,0.52) circle [radius=0.3];

        \end{tikzpicture}
        \subcaption{}
        \label{plot:calibration_cell}
    \end{subfigure}

    \vspace{3mm}

    \begin{subfigure}[c]{0.45\textwidth}
        \centering
		\begin{tikzpicture}
            \node[anchor=south west, inner sep=0] (image) at (0,0)
                {\includegraphics[width=0.7\textwidth]{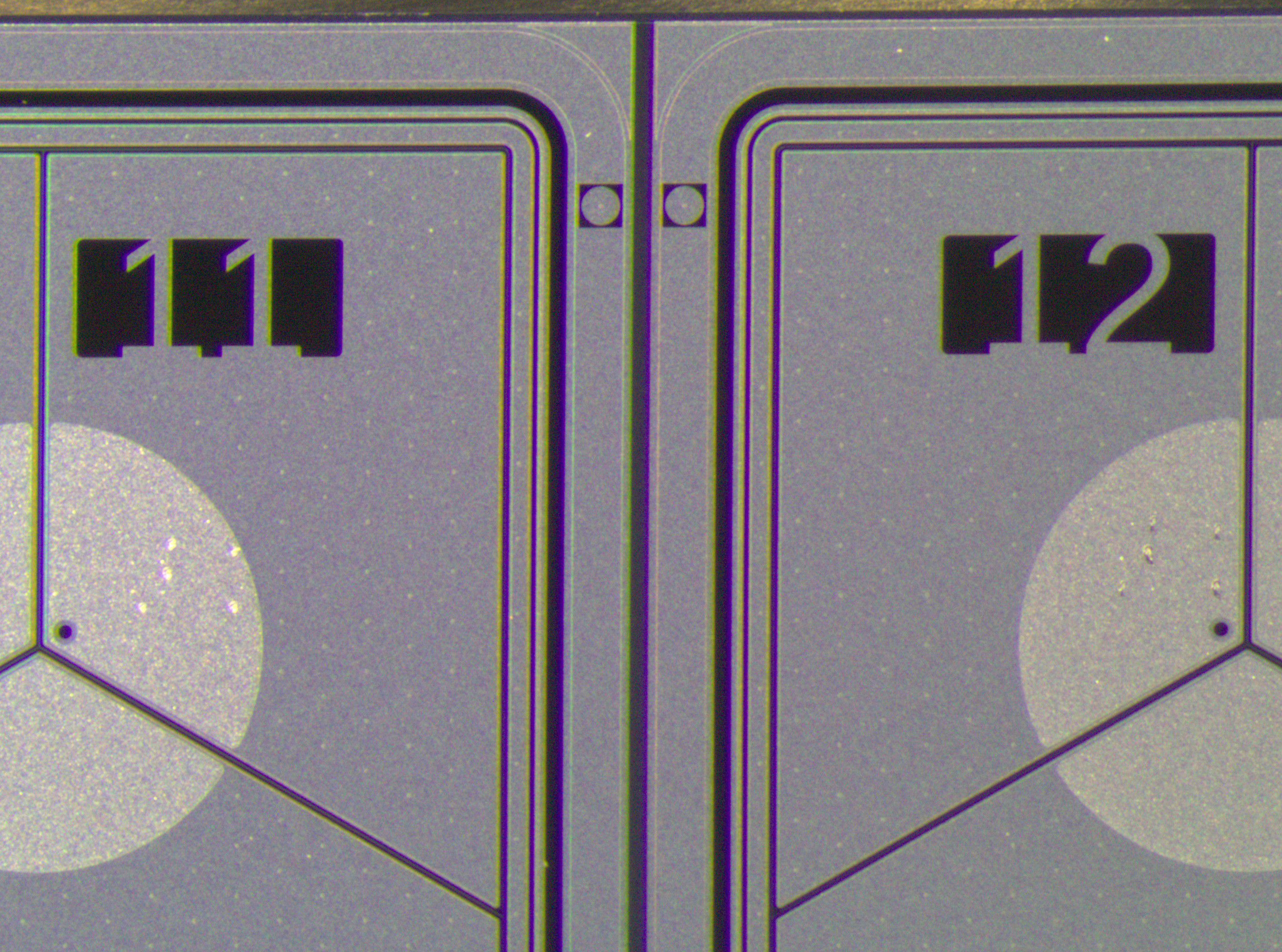}};
			\draw[white, thick] (0.06,0.5) -- ++(0.82,0);
			\node[text=white, anchor=north] at (0.56,0.5) {\footnotesize \SI{1.5}{\milli\metre}};
			\draw[green, thick] (4.35,1.44) circle [radius=0.3];
			\draw[green, thick] (0.75,1.38) circle [radius=0.3];
        \end{tikzpicture}
        \subcaption{}
        \label{plot:edge_cells}
    \end{subfigure}
    \hfill
    \begin{subfigure}[c]{0.45\textwidth}
        \centering

		\begin{tikzpicture}
            \node[anchor=south west, inner sep=0] (image) at (0,0)
                {\includegraphics[width=0.7\textwidth]{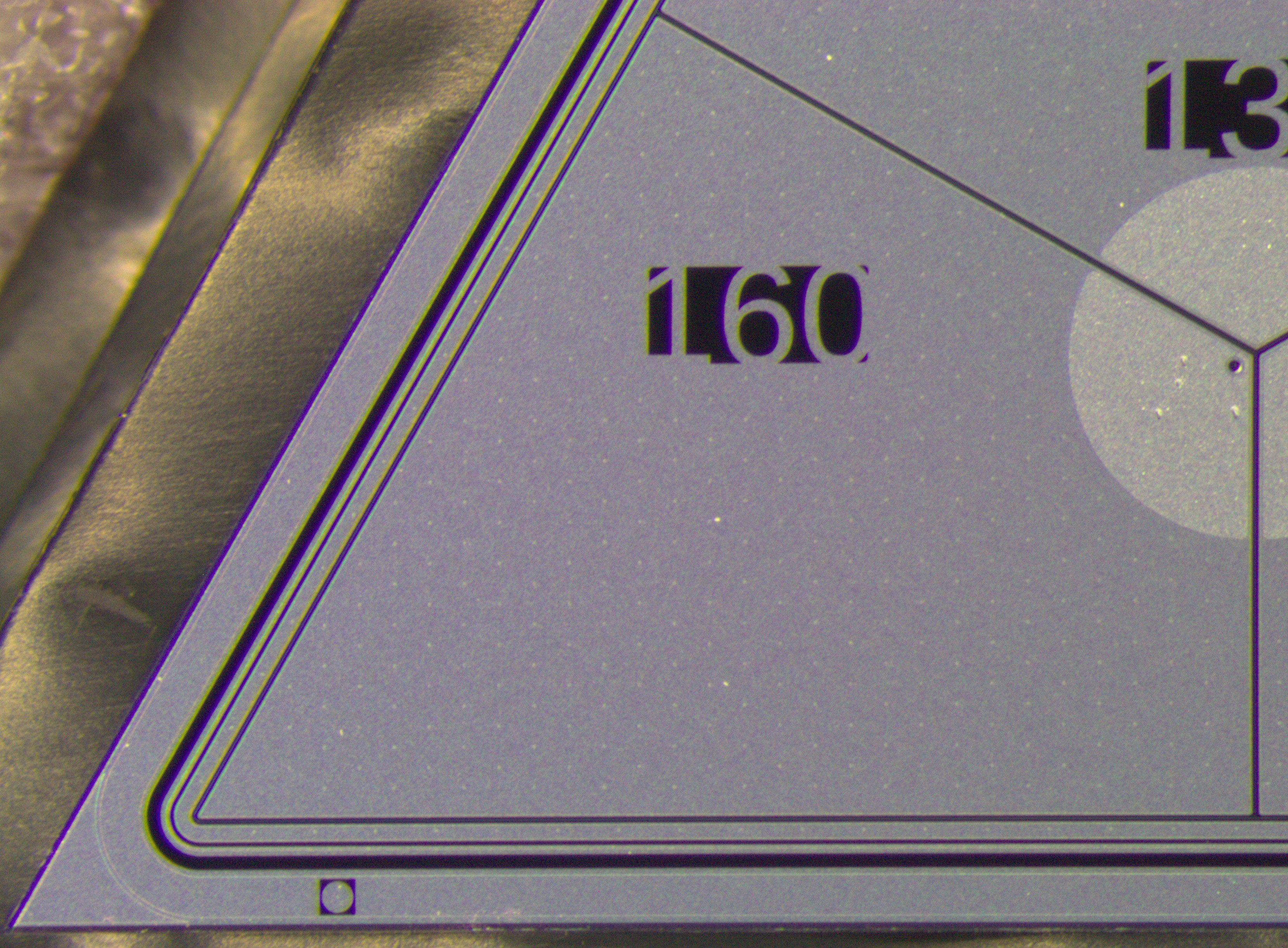}};
			\draw[white, thick] (0.06,0.5) -- ++(0.7,0);
			\node[text=white, anchor=north] at (0.56,0.5) {\footnotesize \SI{1.5}{\milli\metre}};
			\draw[green, thick] (4.45,2.1) circle [radius=0.3];
        \end{tikzpicture}
        \subcaption{}
        \label{plot:edge_cell}
    \end{subfigure}

    \caption{
        Microscope images of various cell types on an HD Top sensor: 
        a) standard full cell, b) calibration cell, c) edge cells near internal dicing lines, and d) edge cell at outer perimeter. 
		The visible contact marks from previous measurements are highlighted with green circles.
		They lie within the openings of the sensor passivation layer (visible as brighter regions on the sensor surface), ensuring proper electrical contact.
		Their positions also enable an estimate of the precision and repeatability of the contacting method.
    }
    \label{plot:cell_types}
\end{figure}

In addition to the standard (full) cells, CE sensors incorporate other specialized cell types, such as edge cells and calibration cells, which differ in area size.
Examples of several cell types are shown in figure~\ref{plot:cell_types} for an HD Top partial sensor.
Edge cells ensure complete area coverage of the detector geometry, while calibration cells with their small surface area maintain a sufficient signal-to-noise ratio with minimum ionizing particles until the End of Life of the detector and will be used for energy calibration.

This study examines whether internal guard rings and HV lines in partial sensors pose a risk of increased leakage currents in the internal sensor cells. 
This is important because the leakage current must remain within the limits imposed by the detector design, the power supply, and the readout chip specifications. 
The total sensor leakage current must stay below \SI{10}{\milli\ampere}~\cite{hgcal-tdr:2018}, and the leakage current per cell must remain below \SI{50}{\micro\ampere} to satisfy the HGCROC3 requirements~\cite{CMS:2022lry}. 
Exceeding these limits can lead to unstable operation and, as observed previously, may even result in sensor damage \cite{Acar_2023}. For this reason, it is necessary to verify whether partial sensors show an increased risk of such behaviour.
\section{Neutron irradiation at the Rhode Island Nuclear Science Center}
\label{sec:Irradiation}

Sensors in this work were neutron-irradiated in the Rhode Island Nuclear 
Science Centre (RINSC). This facility was qualified for the 
irradiation of the CE sensors, as described in ref.~\cite{Acar_2023}.

To position the sensors close to the reactor core, an 8" beam port was used. 
This port is accessible only when the reactor is off, and the samples were typically removed one day after irradiation. 
Since the reactor operated at approximately constant power, the delivered fluence could, to first approximation, 
be controlled by adjusting the irradiation time. 

Due to reactor turn-on effects, a fixed irradiation-time uncertainty of 2 min was assumed. Using the fluence ($\Phi$) scaling factor as described in section~\ref{subsec:Fluence_assessment}, 
this corresponds to an absolute fluence uncertainty of $0.93\times10^{14}~\neqcm$. This contribution was combined in quadrature with a relative uncertainty of \SI{10}{\percent} associated with the reactor power:
\begin{equation}
\label{eq:fluence_error}
\Delta\Phi =\sqrt{(0.10\cdot\Phi)^2+(0.93\times10^{14})^2},
\end{equation}

The resulting absolute fluence uncertainties are $(1.1,\;5.3,\;\mathrm{and}\;13.5)\times10^{14}~\mathrm{n_{eq}/cm^2}$
for irradiation times of 10.9, 113.8, and 292.2 min, respectively.  These correspond to relative fluence uncertainties of approximately \SI{20.9}{\percent}, \SI{10.1}{\percent}, and \SI{10.0}{\percent}.

\begin{figure}
	\centering
    \includegraphics[width=0.9\textwidth]{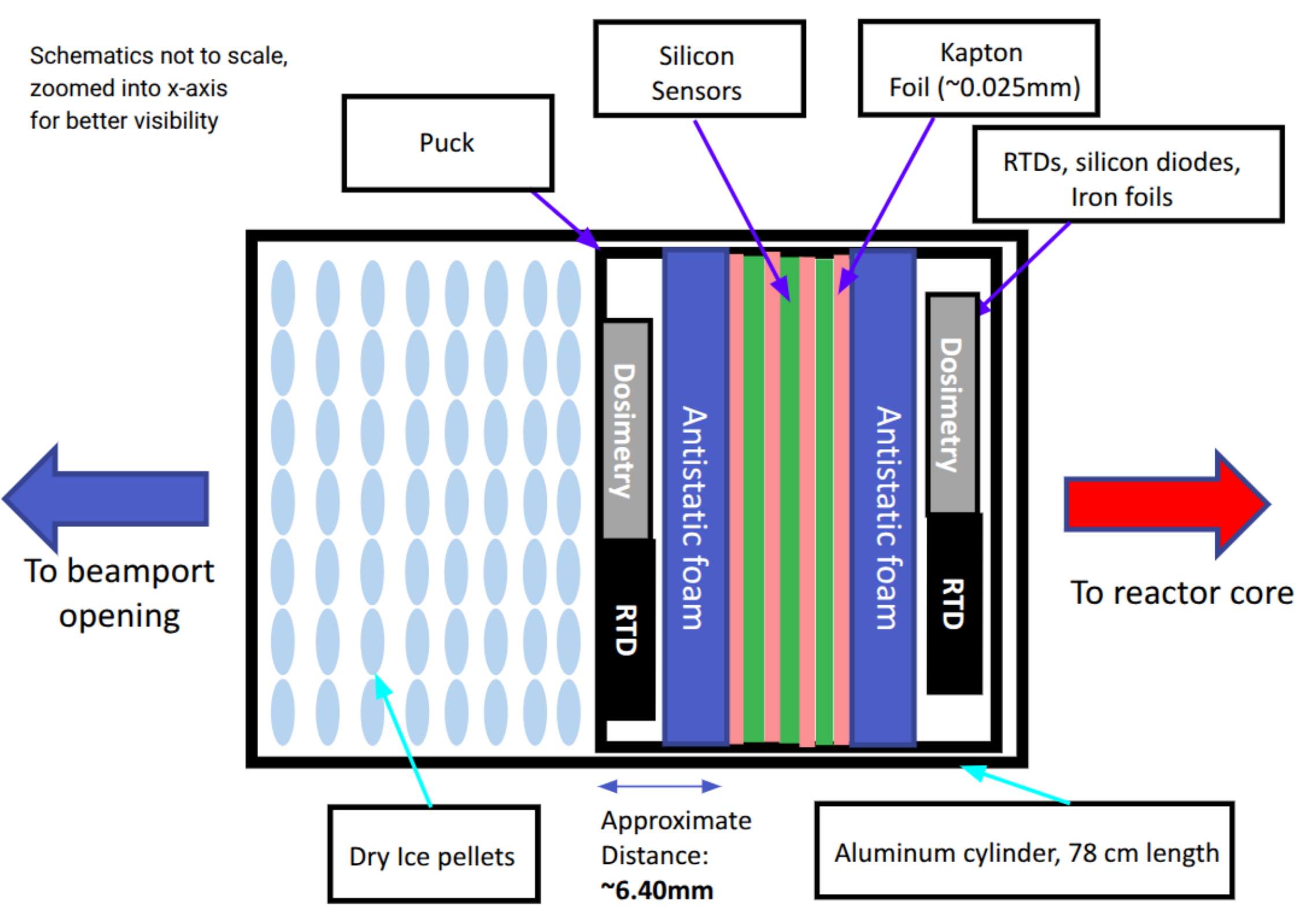}
	\caption{
        Schematics (not to scale) of the sensor placement within the aluminium transport cylinder at RINSC, 
		as described in ref.~\cite{Acar_2023}. The positions of the PT1000 RTDs and dosimetry sensors are 
        presented with the approximate dimensions within the puck 
        container.
		}
	\label{plot:Reactor}
\end{figure}

The sensors are placed in so-called pucks and in aluminium cylinders 
filled with dry ice to limit in-reactor annealing. Schematics shown in figure~\ref{plot:Reactor}
indicate the CE silicon sensors marked in green, the dosimetry sensors in grey, and the temperature sensors in black. The aluminium cylinder is 
filled with dry ice on the side facing the beam port opening.

\subsection{Procedure adaptations for high-fluence irradiations}
\label{subsec:High_fluence_adaptations}
Compared to the study presented in ref.~\cite{Acar_2023}, this study investigated up to 30$~\%$ higher fluences. Achieving higher fluence required keeping the samples in the running reactor for a 
longer duration, which led to increased sample temperatures. This, in turn, introduced undesired 
annealing to the sensors, which had to be carefully monitored and controlled.

Several measures were introduced to mitigate in-reactor annealing, as summarized in table~\ref{table:Cooling_measures}. Details of the puck design, including the ventilation holes, were presented in the previous study~\cite{Acar_2023}. 
Most of these measures were implemented already for the first irradiation round at RINSC. 
Starting from round 9 of the version 2 campaign, high-fluence irradiation rounds (>$1 \times 10^{16}~\neqcm$) 
were split into two parts of equal duration, with dry ice refilled between the two sessions.

The results summarized in table~\ref{tab:summary} indicate that splitting high-fluence rounds is an effective approach to reduce in-reactor annealing.
An exception occurred in round 10, which, although split, experienced over-annealing during the second part due to limited dry ice availability. This was a logistical constraint specific 
to that round and does not contradict the overall effectiveness of the splitting strategy demonstrated in the other high-fluence rounds.

Furthermore, several new puck materials were introduced with the aim of improving the thermal conductivity in high-fluence rounds, 
as shown in table~\ref{tab:summary}. However, no clear effect of the puck material was observed at similar fluence levels.

A discussion on the potential of pre-cooling of the aluminium cylinder is provided in appendix~\ref{appendix:pre-cooling}.

\begin{table}[h]
	\centering
	\caption{The introduced measures to reduce in-reactor annealing.
	}
	\label{table:Cooling_measures}	
	\begin{tabular}{lcc}
		\textbf{Introduced measure} & \textbf{Introduction date} & \textbf{Applied in version, round} \\
		\hline
		Dry ice to cool the cylinder & 26.08.2020 & V1-2: all rounds\\
		Ventilation holes & 26.08.2020 & V1-2: all rounds\\
		High-fluence round splitting & 08.03.2022    & V2: R9, R10, R12, R14    \\
        Pre-cooling of the aluminium cylinder& 18.08.2022 & V2: R14 second part\\
        Puck material optimization & See table~\ref{tab:summary}  & See table~\ref{tab:summary}\\
		\hline
	\end{tabular}
\end{table}

\subsection{Fluence assessment options}
\label{subsec:Fluence_assessment}

The fluence delivered to the samples during irradiation rounds was estimated in 
various ways. As discussed earlier in this section, since the reactor operated at approximately constant power, 
the delivered fluence could be determined by multiplying the irradiation time by a known time-fluence conversion factor.
The irradiation time itself was calculated from the reactor's turn-on/turn-off times, 
which were recorded to the nearest decimal of a minute for all irradiations.

In the version 1 campaign, an initial conversion factor was used to determine the irradiation time needed to achieve the required target fluence. This factor was based on the 
fluence assessment using the diodes that were studied for usage in the D0 experiment~\cite{D0diodes}. However, a comparison with other 
irradiation facilities revealed that the factor overestimated the delivered fluence.

Between the version 1 and version 2 
campaigns the knowledge of the D0 diode thickness was refined: initially assumed to be \SI{240}{\micro\meter}, 
it was later determined to be \SI{290}{\micro\meter}. This discrepancy contributed to the fluence overestimate during the version 1 campaign.
Using the updated thickness, the diode-based fluence is typically about \SI{17}{\percent} lower than previously reported.

Therefore, in the version 2 irradiation campaign, an irradiation time of \SI{21.5}{\min} 
was used to achieve a fluence of $1 \times 10^{15}~\neqcm$. For consistency, 
the conversion factor from the version 1 campaign was retrospectively adjusted in this study to match the value applied in version 2.
Therefore, this change does not contribute to the results.

In addition, various types of fluence dosimetry sensors were placed inside the 
puck, both at the front and back, as shown in figure~\ref{plot:Reactor}. 
The dosimetry sensors used in the version 1 campaign were 
described in ref.~\cite{Acar_2023}, while the version 2 campaign introduced 
additional silicon test structures (diodes) from the CE wafer with an active 
thickness of \SI{120}{\micro\meter}. The depletion voltage measurements of the CE test structures were performed 
at Brown University
to assess the depletion voltage, the associated dark current and ultimately the fluence during the irradiations. 
Since the test structures are also diodes from the same CE wafer as the tested CE sensors, 
this might have introduced a potential correlation between the measurement and the calibration.
 The maximum expected correlation was conservatively 
 estimated to be $\sim\SI{50}{\percent}$ when averaging CE- and D0-based fluence estimates 
 (lower and intermediate fluence range: $< 3\times10^{15}~\neqcm$), and to be absent when using 
 iron-foil activation for the fluence determination (high fluence range).
 However, for the final analysis, the fluence is determined from the irradiation time, 
 as described in section~\ref{subsec:Validation_of_fluence_assessment_procedure}, and therefore this potential correlation does not affect the fluence values used in this study.

Fluence measurements from the respective dosimetry sensors were averaged, 
following the procedure described in ref.~\cite{Acar_2023}. It is relevant to mention 
that the location of the sensors within the puck influences the results of the 
fluence measurement, due to the varying distance from the dry 
ice, as well as due to the fluence profile as discussed 
in section~\ref{sec:LeakageCurrent}. 
To quantify the scale of position-dependent fluence variations within the puck, the relative median absolute deviation (RMAD = MAD/median) of the fluence values measured across sensor locations was used.
After applying the procedure adaptations for high-fluence irradiations described in section~\ref{subsec:High_fluence_adaptations}, the RMAD remained below \SI{8}{\percent} for all campaigns that were not annealed 
into the reverse-annealing regime (compare table~\ref{tab:summary}). Overall, the RMAD exceeded \SI{10}{\percent} only for campaigns 7 and 9 of version 1, reaching \SI{11}{\percent} and \SI{17}{\percent} respectively.

\subsection{In-reactor annealing assessment}
\label{subsec:Annealing_assessment}
For the temperature profile measurement during the irradiation rounds, RTDs 
were placed inside the puck, as can be seen in figure~\ref{plot:Reactor}. 
There were two 
sensors placed in the front and one in the back, yet for some rounds 
there were fewer sensors available due to system failures. If available, the estimation of the temperature experienced by 
the silicon sensors in the middle of the puck was calculated as an average between the back and front RTD readouts.
Starting from round 7 of the version 1 campaign, additional front- and back-side RTDs were introduced, 
improving the determination of the average sensor temperature and, consequently, the assessment of the 
equivalent in-reactor annealing time.

\begin{figure}
	\centering
	\includegraphics[width=0.8\textwidth]{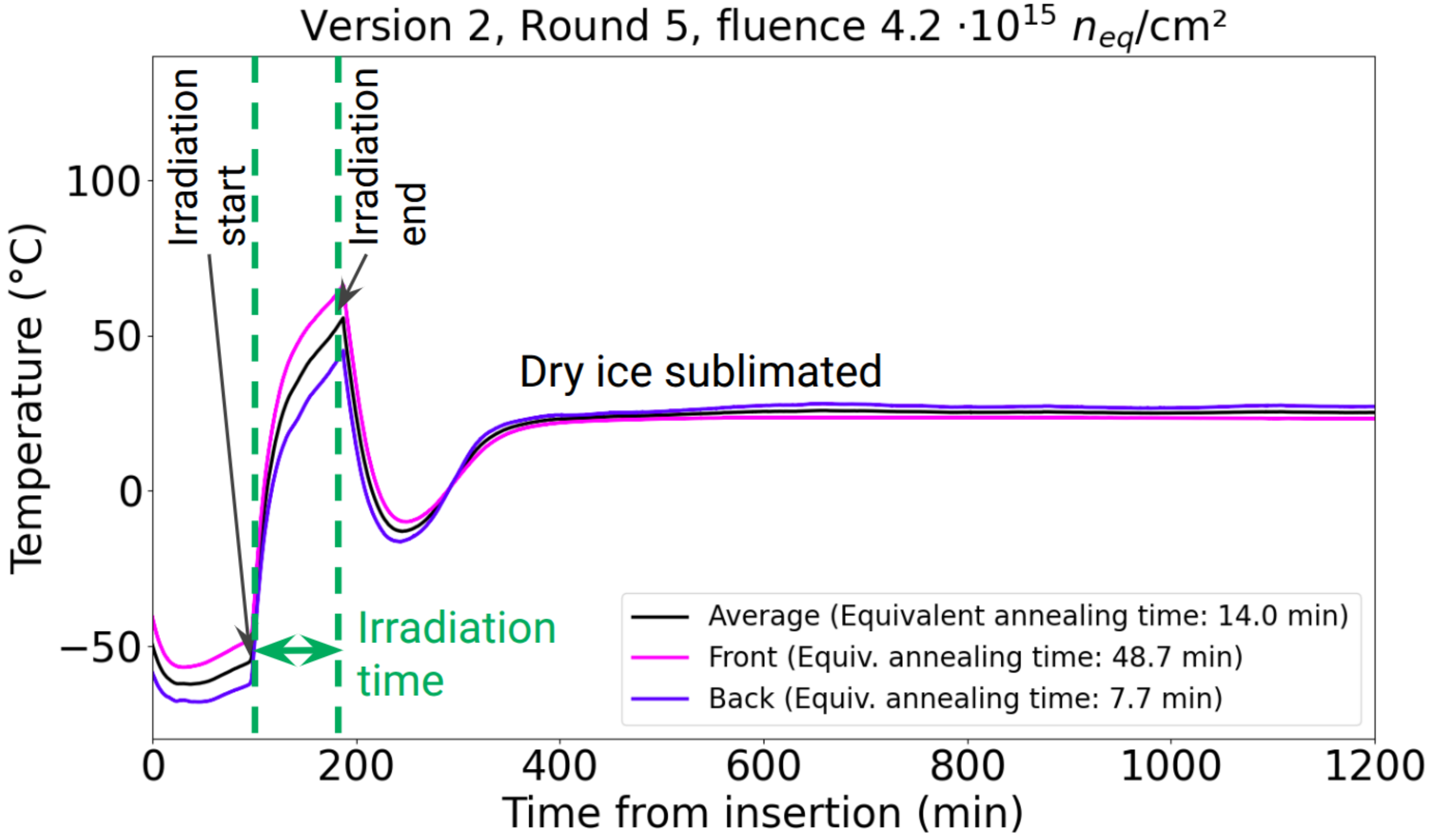}
	\caption{
		An example of a temperature profile obtained from an irradiation round at RINSC, showing averaged measurements at the front,
		 back, and the average of both sides. The distinct phases of irradiation can be inferred from the shape of the profile.
	}
			
	\label{plot:Temp_profile}
\end{figure}

An example of a complete temperature profile recording from an irradiation round is shown in figure~\ref{plot:Temp_profile}.
The equivalent annealing time at \SI{60}{\celsius} was calculated from the average temperature profile using the so-called Hamburg model, as described in ref.~\cite{Moll:300958}.
This parametrization is widely used across different fluence levels; however, it was originally developed for n-type sensors and does not directly apply to the p-type sensors foreseen for the CE upgrade.

An updated annealing model, validated for the FZ and epitaxial silicon materials used in CE sensors over the relevant fluence range,
has recently been published~\cite{Diehl2026}. To estimate the systematic uncertainty associated with the use of the Hamburg model,
the equivalent annealing times obtained with the two parametrizations were compared for all irradiation rounds included in the determination
of the current-related damage constant.

The largest absolute difference between the two parametrizations was observed for the epitaxial sensor irradiation in version~2, round~14, part~1.
For this round, the Hamburg model yields an average equivalent annealing time of \SI{30.2}{\minute},
whereas the updated model yields \SI{66.1}{\minute}, corresponding to an absolute difference of \SI{35.9}{\minute}.

The next-largest difference is \SI{29.3}{\minute} for version~2, round~14, part~2, followed by \SI{24.3}{\minute} for version~2, round~13.
All other FZ irradiations show substantially smaller differences of approximately \SIrange{8}{10}{\minute}.

\begin{figure}
	\centering
	\includegraphics[width=0.8\textwidth]{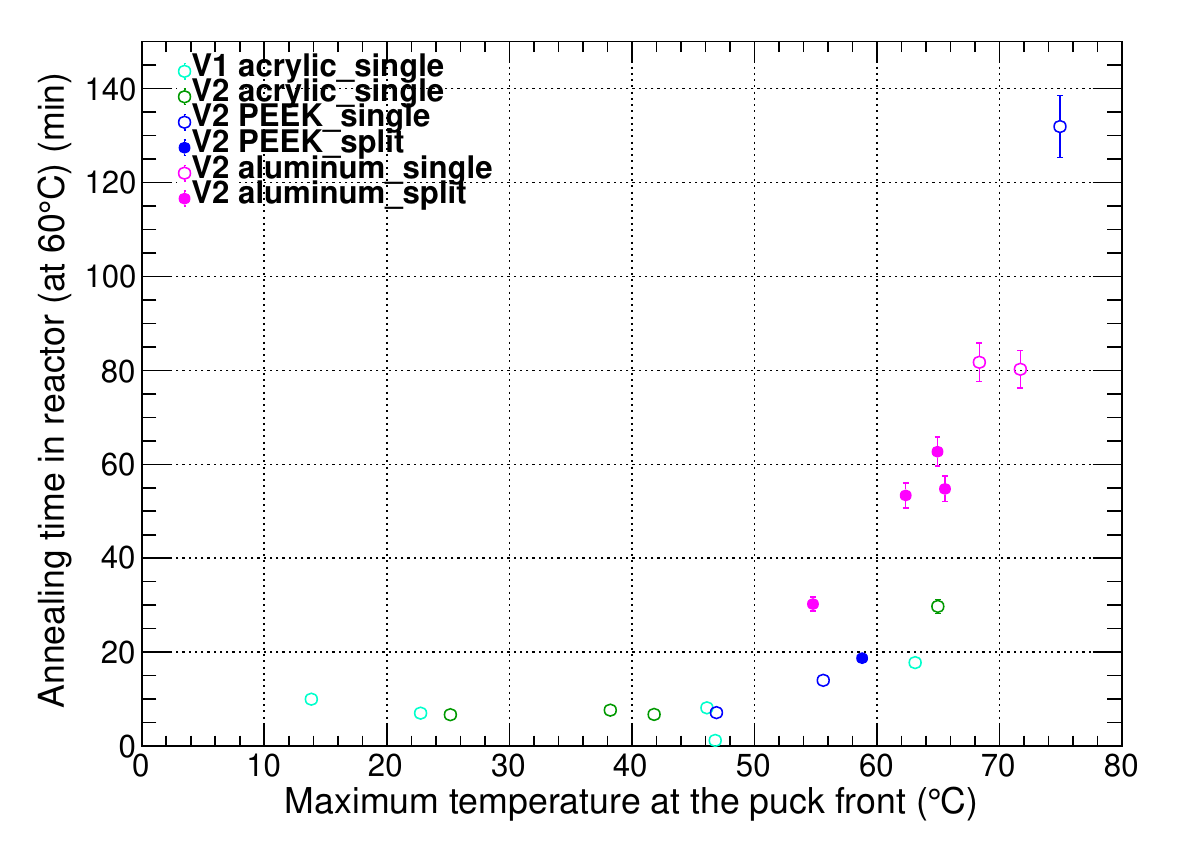}
	\caption{
		Annealing time at the sensor position in the reactor versus the maximum temperature measured at the front of the puck during irradiation.
	}
			
	\label{plot:AnnealingvsTemperature}
\end{figure}

 The equivalent annealing time was strongly impacted by the short time at 
 high temperature during the irradiation, as illustrated in figure~\ref{plot:AnnealingvsTemperature}. Additionally, because the calculation included waiting times of up 
 to one day post-irradiation, it was also influenced by extended periods at room temperature after the dry ice has sublimated, 
 which could vary depending on the season of the irradiation rounds. 
 The average in-reactor annealing time increased with fluence for rounds irradiated in a single part, as expected.

The chosen in-reactor annealing measurement method introduced uncertainties in estimating the equivalent annealing time for the CE silicon sensors.
These uncertainties were caused by missing back-side measurements in some rounds and significant discrepancies between the front and back temperature readings in others, as shown in table~\ref{tab:summary}.
The impact of these discrepancies has been previously studied using leakage current versus annealing time measurements for CE diodes, as discussed in ref.~\cite{Kieseler_2023}.

\section{Leakage current profiles across sensors}
\label{sec:LeakageCurrent}

\begin{figure}
	\captionsetup[subfigure]{aboveskip=-1pt,belowskip=-1pt}
	\centering
    \includegraphics[width=0.4\textwidth]{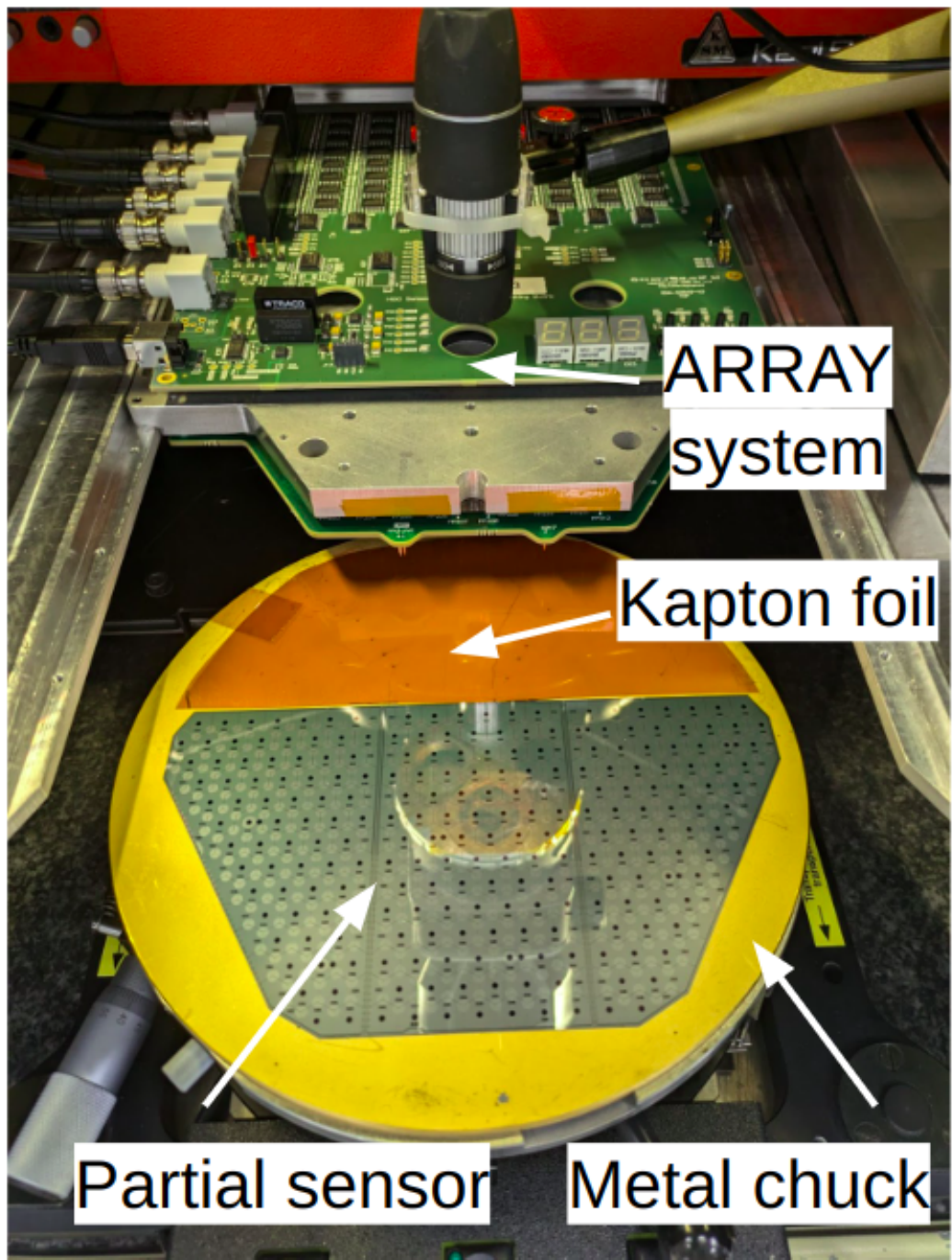}
	\caption{
        A low-density CE silicon partial sensor of type LD Five 
        before connecting to the switch- and probe-card of the non-temperature 
        controlled ARRAY System at CERN. The missing sensor area is covered 
        with Kapton foil. The arrangement is equivalent to the one utilized 
        in the measurements of the irradiated sensors.
		}
	\label{plot:Partial_Measurement}
\end{figure}

The system to perform the electrical characterization of the irradiated sensors as 
well as the characterization procedure was described in detail in ref.~\cite{Acar_2023}. 
The measurement was done using a probe- and switch-card system called ARRAY~\cite{pitters:array2019}. New probe cards were designed and fabricated for partial LD and HD multigeometry wafers. They have been tested on 
non-irradiated partial sensors first. Pins of the probe cards that were 
not used for the partial sensor type measured were left open by contacting 
them to Kapton foil, placed at the probe-station chuck next to the partial 
sensor, as can be seen in figure~\ref{plot:Partial_Measurement}. Except for the measurement discussed in section~\ref{sec:Temperature}, all sensor measurements were performed on a chuck set to \SI{-40}{\celsius}. The chuck 
(C200-40 model, manufactured by ATT Systems) is specified by the manufacturer to have a temperature accuracy and stability of $\SI{\pm0.1}{\celsius}$~\cite{att-chucks}. 
The horizontal temperature uniformity of the chuck surface was taken into account following the procedure described in ref.~\cite{Acar_2023}.
The typical uncertainty of the voltage measurements is below 2$~\%$, while the uncertainty of the leakage current is below 0.1$~\%$. In all plots in this study,
 these uncertainties are smaller than the symbol size.

\subsection{Voltage dependence of leakage current profiles}
\label{subsec:Voltage_Dependence_of_Leakage_Current-Profiles}

\begin{figure}
	\captionsetup[subfigure]{aboveskip=-1pt,belowskip=-1pt}
	\centering
    \includegraphics[width=\textwidth]{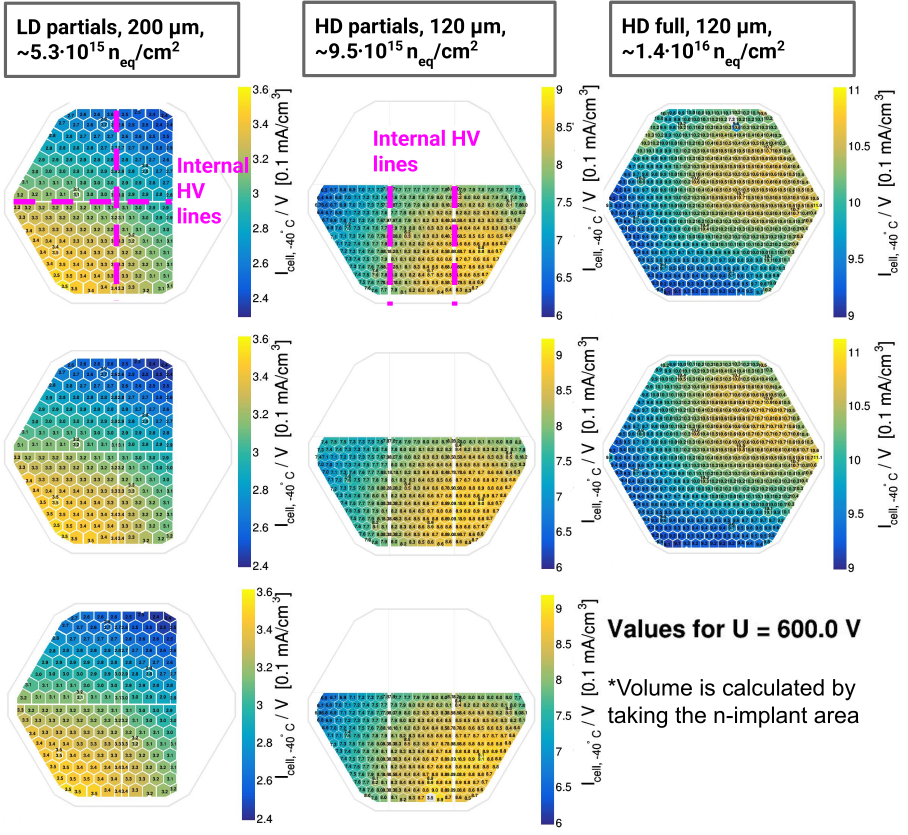}
	\caption{
        Per-cell volume-normalized leakage currents, interpolated to an effective bias voltage of 
        600 V, for three representative irradiation rounds. 
		The first and second columns show profiles for partial sensors. 
		}
	\label{plot:per_cell_scaled}
\end{figure}

Figure~\ref{plot:per_cell_scaled} presents the per-cell leakage currents, interpolated to an effective bias voltage of 600V. 
The cell leakage current was normalized by the cell volume, where the n-implant area was used to define the cell area. 
As examples, results from two irradiation rounds with partial sensors and one high-fluence round with a full sensor were selected.
These results confirm the presence of similar leakage current profiles across the surfaces of sensors from the same irradiation round, 
consistent with the observations reported in ref.~\cite{Acar_2023}.

The leakage current profiles could have two sources: a fluence gradient within the reactor, or an annealing time 
variation caused by temperature differences across the sensor, resulting from uneven cooling during the irradiation.

\begin{figure}
	\captionsetup[subfigure]{aboveskip=-1pt,belowskip=-1pt}
	\centering
    \includegraphics[width=0.8\textwidth]{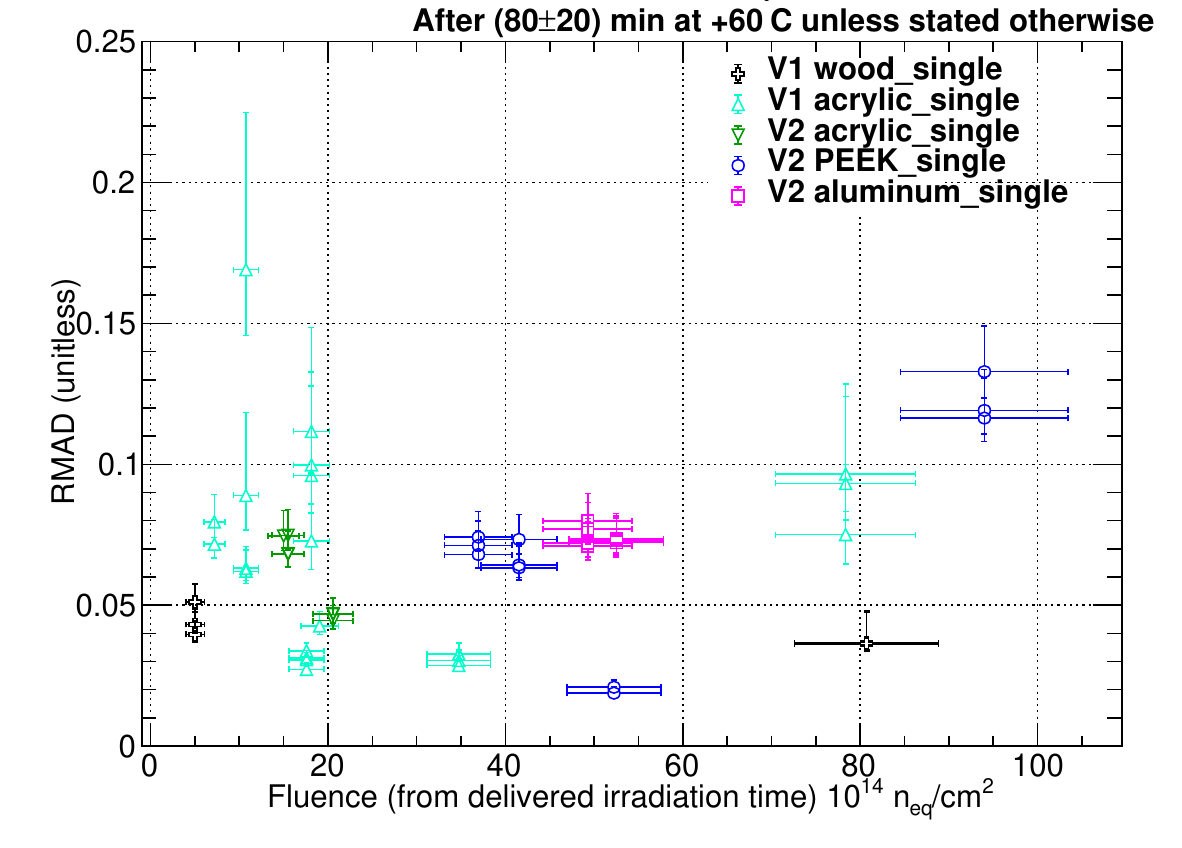}
	\caption{
        Leakage current dispersion (RMAD, compare section~\ref{subsec:Fluence_assessment}) versus fluence for different puck materials.
		}
	\label{plot:RMADvsFluence}
\end{figure}

To investigate the origin more closely, the leakage current distribution across the sensor was analysed, 
taking into account the differences between puck materials and the effects of split versus single rounds (see section~\ref{subsec:High_fluence_adaptations}). 
This is motivated by the fact that the puck materials differ 
in their thermal conductivity, which could lead to variations in the cooling 
homogeneity.
The analysis shows no clear correlation between the puck material and the leakage current distribution across the sensors, as shown in figure~\ref{plot:RMADvsFluence}.
This suggests that the observed leakage current profiles are likely due to the fluence profile in the reactor rather than the annealing profile. 

\begin{figure}
	\captionsetup[subfigure]{aboveskip=-1pt,belowskip=-1pt}
	\centering
	\begin{subfigure}[b]{0.49\textwidth}
		\begin{tikzpicture}
            \node[anchor=south west,inner sep=0] (image1) at (0,0) {\includegraphics[width=1.0\textwidth]{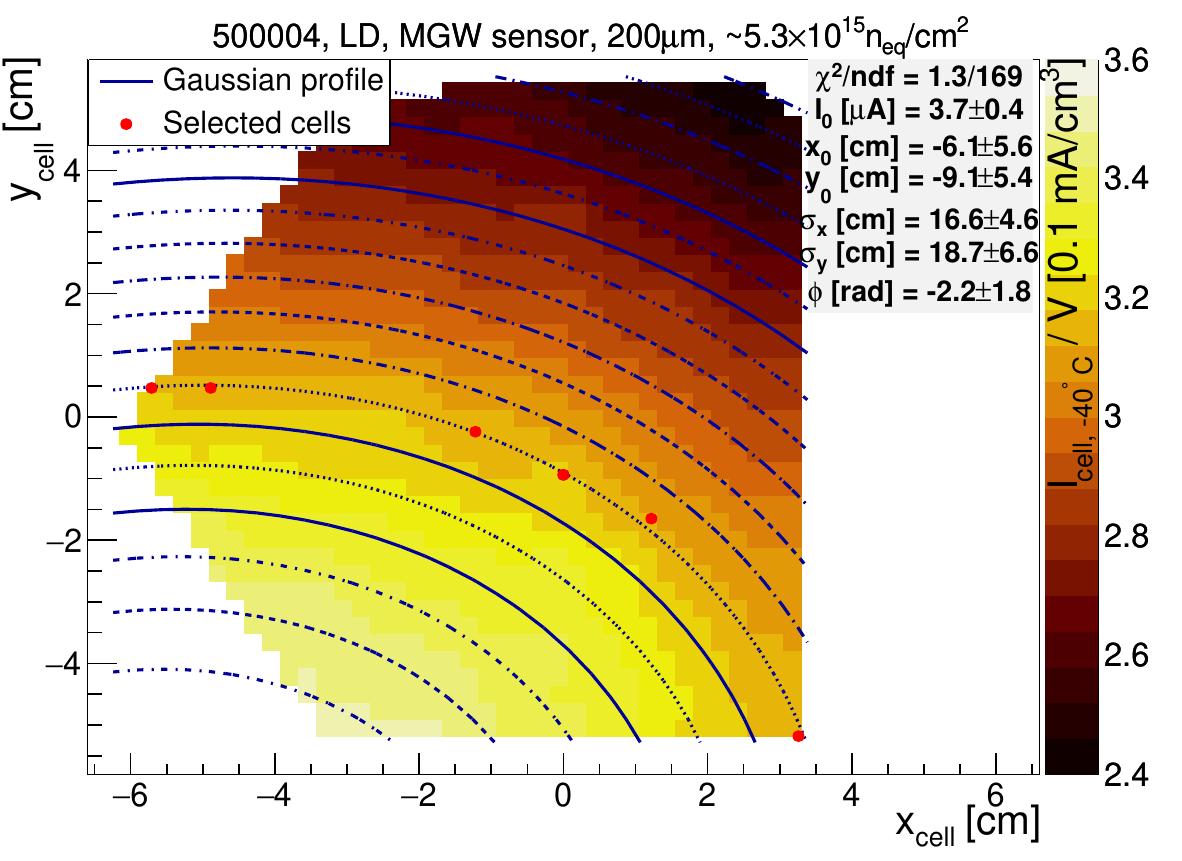}};
        \end{tikzpicture}
		\subcaption{
		}
        \label{plot:LD_Partial_Hexplot}
	\end{subfigure}
	\begin{subfigure}[b]{0.49\textwidth}
		\begin{tikzpicture}
            \node[anchor=south west,inner sep=0] (image1) at (0,0) {\includegraphics[width=1.0\textwidth]{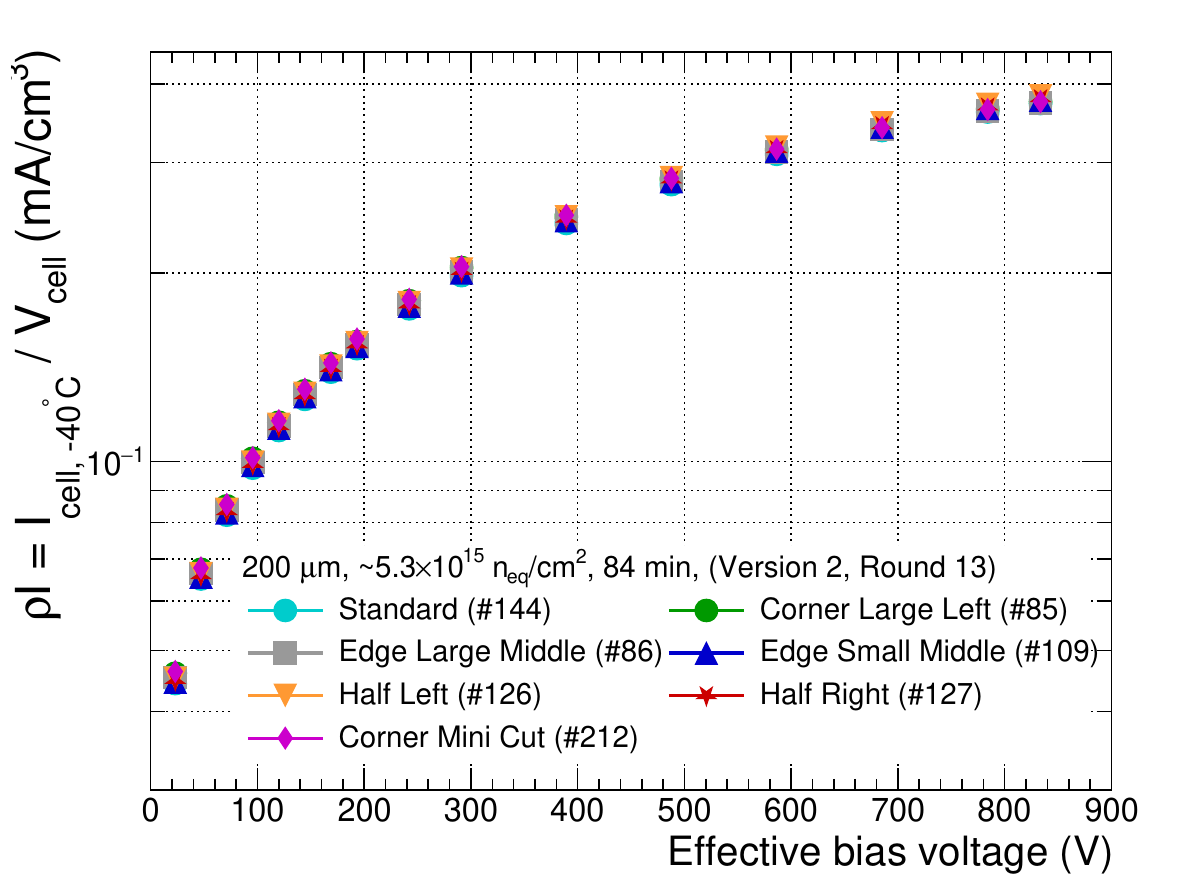}};
            \begin{scope}[x={(image1.south east)},y={(image1.north west)}]
				\node[anchor=south east] at (0.53,0.8) {LD Partial (Five)};
            \end{scope}
        \end{tikzpicture}
		\subcaption{
		}
        \label{plot:LD_Partial_not_scaled}
	\end{subfigure}
	\begin{subfigure}[b]{0.49\textwidth}
		\begin{tikzpicture}
            \node[anchor=south west,inner sep=0] (image1) at (0,0) {\includegraphics[width=\textwidth]{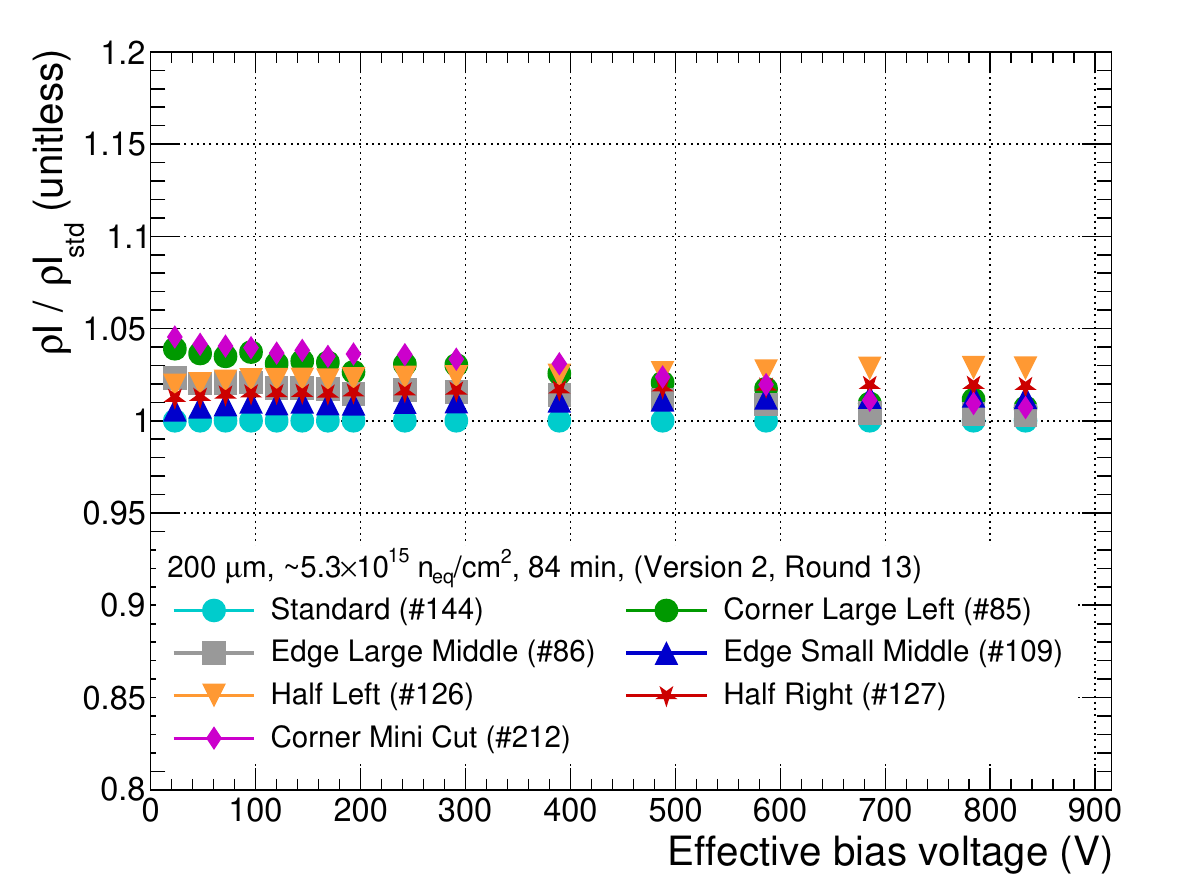}};
            \begin{scope}[x={(image1.south east)},y={(image1.north west)}]
				\node[anchor=south east] at (0.53,0.8) {LD Partial (Five)};
            \end{scope}
        \end{tikzpicture}
		\subcaption{
		}
        \label{plot:LD_Partial}
	\end{subfigure}
	\hfill
	\begin{subfigure}[b]{0.49\textwidth}
		\begin{tikzpicture}
            \node[anchor=south west,inner sep=0] (image1) at (0,0) {\includegraphics[width=\textwidth]{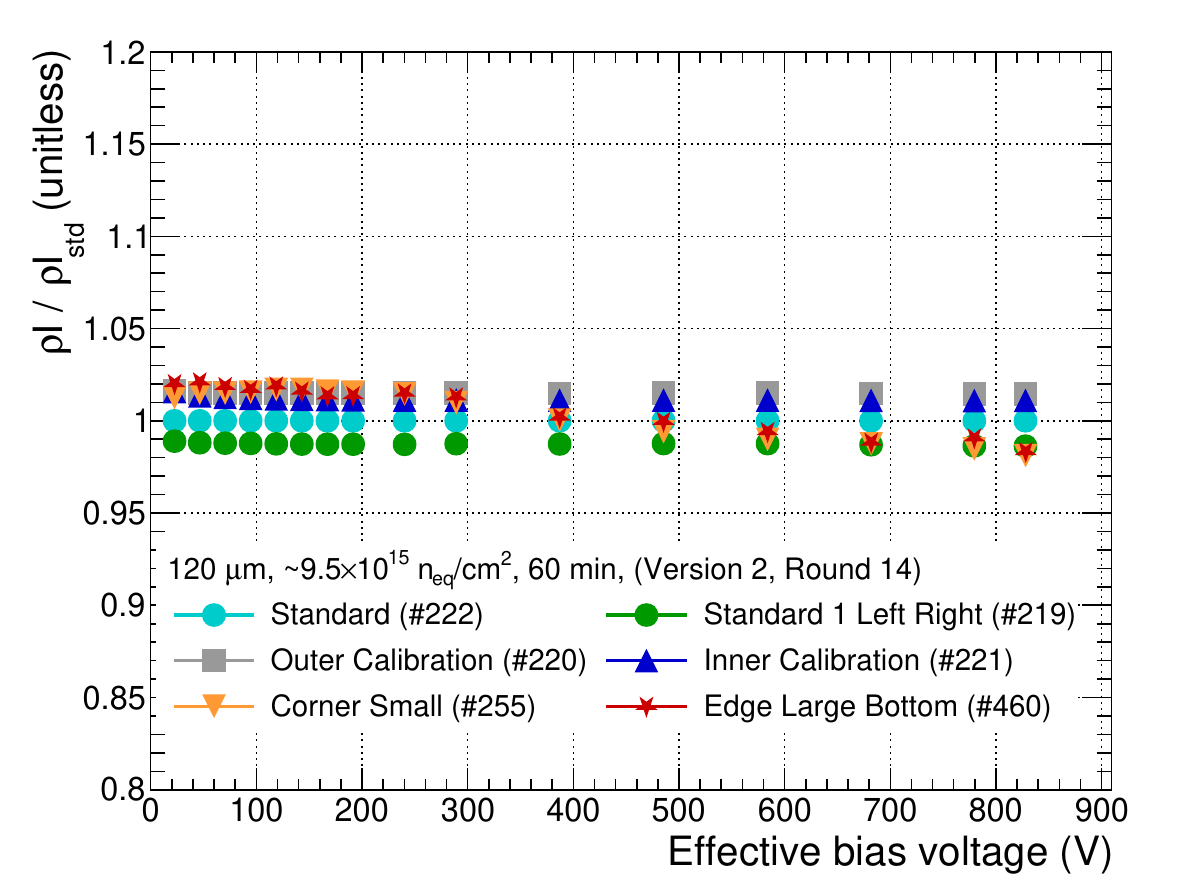}};
            \begin{scope}[x={(image1.south east)},y={(image1.north west)}]
				\node[anchor=south east] at (0.6,0.8) {HD Partial (Bottom)};
            \end{scope}
        \end{tikzpicture}
		\subcaption{
		}
        \label{plot:HD_Partial}
	\end{subfigure}
	\begin{subfigure}[b]{0.49\textwidth}
		\begin{tikzpicture}
            \node[anchor=south west,inner sep=0] (image1) at (0,0) {\includegraphics[width=\textwidth]{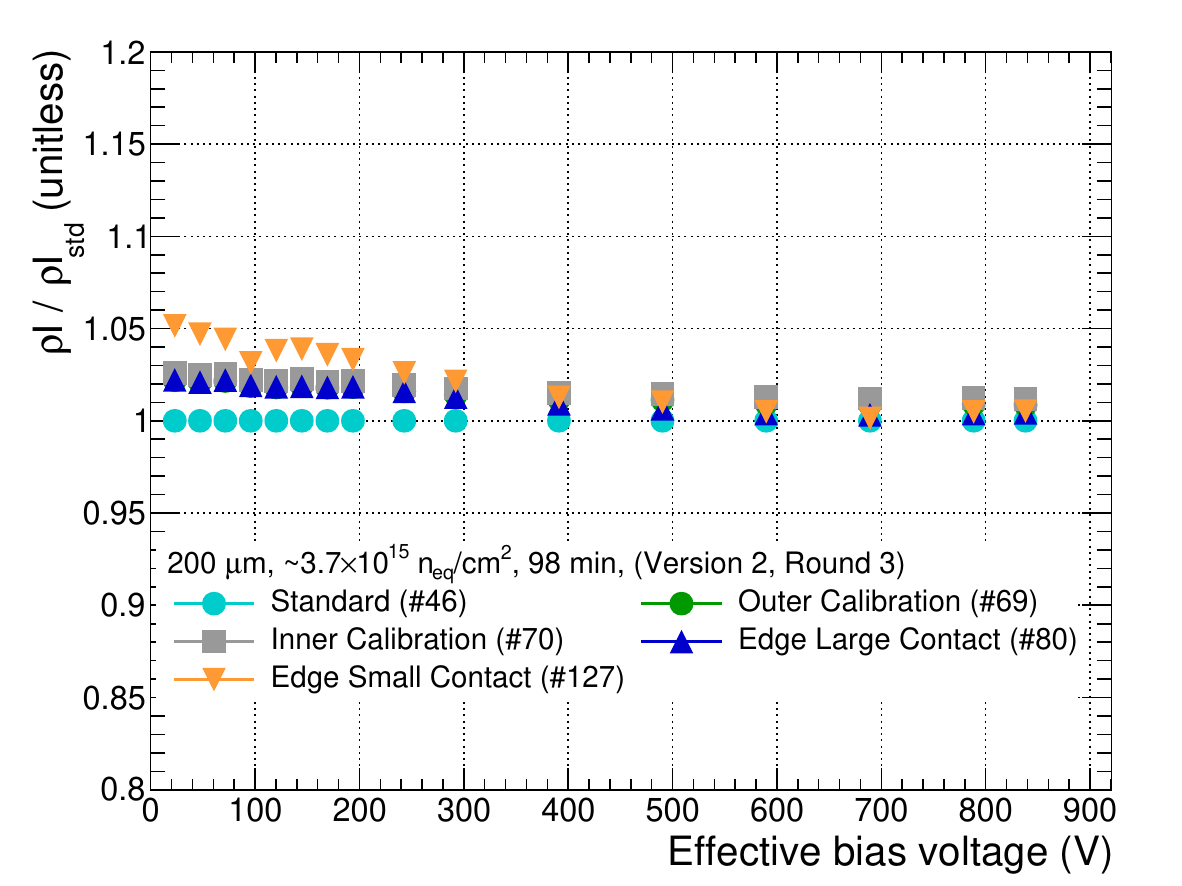}};
            \begin{scope}[x={(image1.south east)},y={(image1.north west)}]
				\node[anchor=south east] at (0.35,0.8) {LD Full};
            \end{scope}
        \end{tikzpicture}
		\subcaption{
		}
        \label{plot:LD_Full}
	\end{subfigure}
	\begin{subfigure}[b]{0.49\textwidth}
		\begin{tikzpicture}
            \node[anchor=south west,inner sep=0] (image1) at (0,0) {\includegraphics[width=\textwidth]{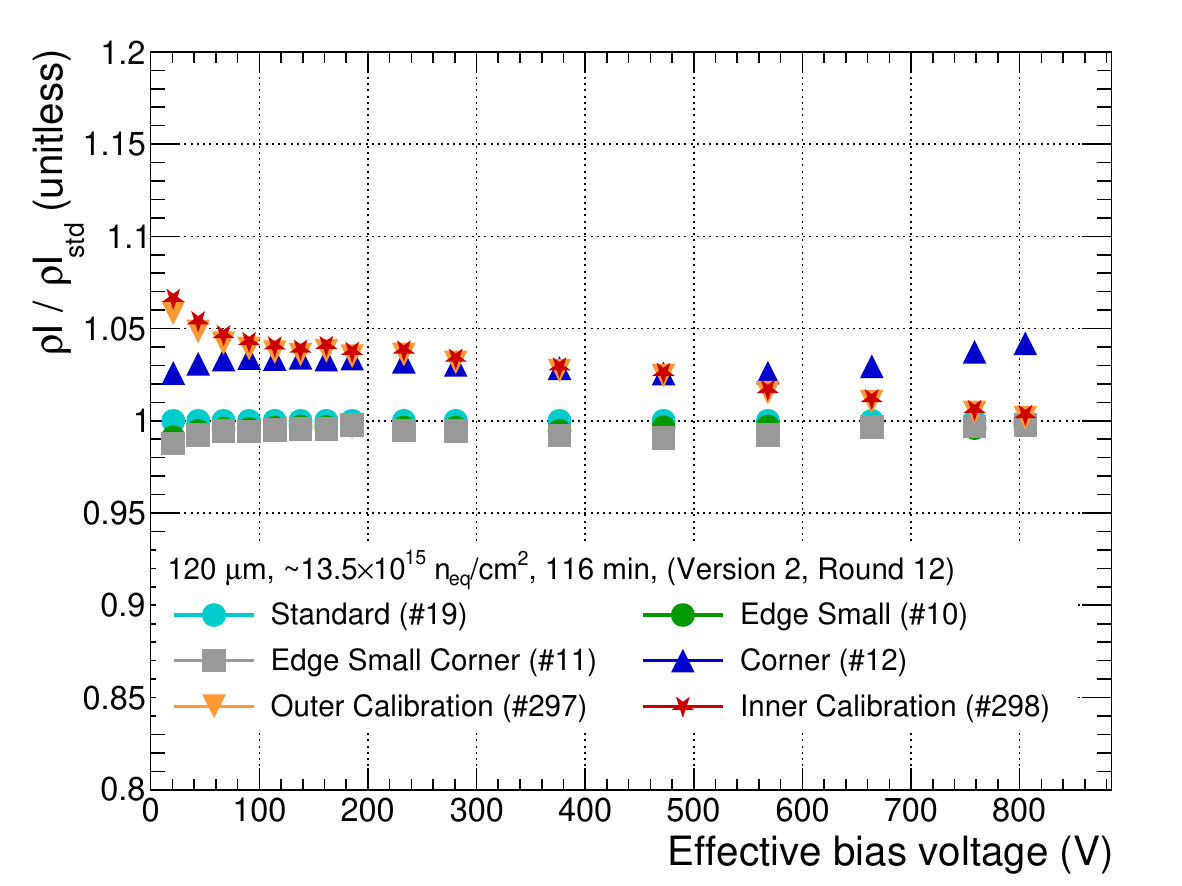}};
            \begin{scope}[x={(image1.south east)},y={(image1.north west)}]
				\node[anchor=south east] at (0.35,0.8) {HD Full};
            \end{scope}
        \end{tikzpicture}
		\subcaption{
		}
        \label{plot:HD_Full}
	\end{subfigure}
	\caption{
		a) Volume-normalized per-cell leakage current distribution displayed as a 2D heatmap at an effective bias voltage of 
		600 V for a representative partial hexagonal LD sensor after 84 minutes of annealing. The plot includes 2D Gaussian fits of the leakage current distribution, 
		with red dots marking selected cells of different shapes aligned along an iso-fluence line. The coordinate system is referenced to the centre of a full sensor.
		b) IV curves for selected cells indicated in (a) for the partial LD sensor.
		 c) Normalized IV curves for cells of different shapes for the LD partial sensor shown in (a),
		 d) for an HD partial sensor, e) for an LD full sensor, and f) for an HD full sensor. The normalization is performed with respect to the leakage current of a standard-shaped cell at each voltage point. 
		All leakage currents are scaled to the respective cell volume.
	}

\end{figure}

The spatial distribution of the leakage current $I(x, y)$ was modelled using a two-dimensional elliptical Gaussian function with rotation,
 defined over the sensor cell coordinates 
$x_{cell}$ and $y_{cell}$, as illustrated in figure~\ref{plot:LD_Partial_Hexplot}. The function takes the form:
\begin{equation} \label{eq:leakage} I(x_{cell}, y_{cell}) = I_0 \cdot \exp\left(-\left[a(x_{cell} - x_{0})^2 + 2b(x_{cell} - x_{0})(y_{cell} - y_{0}) + c(y_{cell} - y_{0})^2\right]\right), \end{equation}
where ($x_{0}$, $y_{0}$) is the centre of the distribution, $I_0$ is the amplitude, $\sigma_x$ and $\sigma_y$ are the widths along the principal axes, and $\phi$ is the rotation angle of the ellipse.
The parameters $a$, $b$, $c$ are derived from $\sigma_x$, $\sigma_y$ and $\phi$.
A fit to the measured data yielded a reduced $\chi^2$ value close to 1. This supports the assumption of a smooth, continuous variation of the leakage current, 
which was used to define "iso-fluence/annealing" contours across the sensor surface, indicated by the blue lines in figure~\ref{plot:LD_Partial_Hexplot}. These "iso-fluence/annealing" lines represent constant leakage current level on the sensor,
and they indicate hypothetical contours across the sensor area where cells experience comparable fluence or annealing conditions, inferred from similar leakage current values. 

To numerically verify whether different cell types across the sensors exhibit similar leakage current behaviour with increasing voltage, 
the volume-normalized IV curves were compared for cells from both full and partial sensors. 
The comparison was performed using cells positioned approximately along the "iso-fluence/annealing" lines, as illustrated in figure~\ref{plot:LD_Partial_Hexplot}). 
Where applicable, cells of different geometries were chosen for the comparison, as shown in figure~\ref{plot:LD_Partial_not_scaled}, to further test whether sensor layout affects current behaviour.

\subsection{Discussion of results for partial sensors}
\label{subsec:Partial_profiles}

As shown in figure~\ref{plot:per_cell_scaled}, the leakage current profiles are smooth for both the partial and full sensors at high fluence.
In particular, for the partial sensors (first two columns), no instabilities related to the sensor layouts
or the cell vicinity to the internal HV protection structures are observed (see section~\ref{sec:Sensors}).
Moreover, it can be observed that the leakage current maximum occurs within the sensor area, 
not in the cut-out region. This is relevant because the cells at maximum leakage current are used in the analysis in sections~\ref{sec:Volume_normalized_current} and \ref{sec:Current_related_damage_rate}
for both partial and full sensors. This observation supports the use of partial sensors for that analysis.

For quantitative comparison, the volume-normalized per-cell leakage current ($\rho I$) was further normalized at each voltage point to the leakage current of a standard-shaped reference cell 
($\rho I_{\text{std}}$). As shown in figures~\ref{plot:LD_Partial}~to~\ref{plot:HD_Full}, the IV curves agree within 
8$~\%$ across different cell types and sensor variants. Moreover, the current variation among differently shaped cells remains stable as the bias voltage increases, 
with the largest observed deviation being 7$~\%$ for an HD full sensor (see figure~\ref{plot:HD_Full}).

This result demonstrates that differently shaped cells, including those near the guard ring for full 
sensors and the internal guard ring for partial sensors, exhibit stable leakage current behaviour with 
increasing bias voltage. This stability, regardless of cell geometry or proximity to HV protection structures, 
proves the robustness of the sensor design and its uniform response to applied bias voltage.
\section{Leakage current dependence on voltage}
\label{sec:Volume_normalized_current}

\subsection{Results on leakage current behaviour}
\label{subsec:Results_CurrentvsVoltage}

In this study, the term "current maximum" refers to the average leakage current of the three full cells showing the highest leakage current within each 
sensor. This region corresponds to the area of maximum neutron fluence, and the resulting value is denoted as 
$I_{\text{cell at fluence max}}$ in the plots.

Most sensors exhibited a typical diode-like leakage current-voltage (IV) dependence, as shown by the coloured graphs in figure~\ref{plot:IV_cell}. 
Overall, 31 of the 36 measured sensors displayed stable leakage current profiles. However, during the initial irradiation rounds with fluences exceeding $1 \times 10^{16}\neqcm$, extreme conditions were reached,
 with temperatures as high as \SI{125}{\celsius}. These conditions resulted in significant in-reactor annealing, 
 equivalent to approximately 10000 minutes at \SI{60}{\celsius}. One example is round 4 of the version 2 campaign, listed in table~\ref{tab:summary}.

Sensors exposed to such prolonged annealing (far into the reverse annealing regime~\cite{moll:SiDamages,MOLL199987}) and high fluences displayed a different IV behaviour. Specifically, the leakage current showed an exponential rise, as shown by the grey graphs in figure~\ref{plot:IV_cell}. 
This effect is discussed in more details in section~\ref{subsec:Analysis_the_Exponential}.

The total current measured by the power supply, when scaled by sensor volume, aligns well with the cell currents in terms of magnitude (see figure~\ref{plot:IV_total}). After scaling, IV curves cluster together for sensors of the same thickness, 
suggesting that the leakage current is predominantly influenced by bulk effects rather than surface phenomena.

\subsection{Analysis of exponential increase in leakage current}
\label{subsec:Analysis_the_Exponential}

The exponential rise in leakage current observed in some highly irradiated sensors after prolonged in-reactor annealing suggests the presence of a field-dependent current enhancement mechanism.
The ARRAY probe-card system used for 8"-sensor characterization does not allow charge collection measurements with external particle or
laser sources. Therefore, a direct experimental investigation of the microscopic origin of the exponential current increase on the full sensors was beyond the scope of this work.

Dedicated diode test structures were included in the wafer layout, specifically to characterize bulk radiation-damage properties that cannot be measured directly on the full sensors~\cite{Kaluzinska2025, Diehl2026}.
The diode test structures were fabricated on the same wafers as the 8" sensors and therefore share the same active thickness, silicon material, and 
fabrication process.
Complementary measurements on these diode test structures showed a simultaneous increase in leakage current and charge collection after prolonged annealing, consistent with charge multiplication~\cite{Diehl2026}.
These measurements therefore provide indirect experimental support for charge multiplication as one possible explanation for the exponential leakage-current increase observed in the full sensors.

For CE sensors, the high-fluence irradiation rounds in which the irradiation was divided into two parts to limit the cumulative in-reactor
annealing time, including version~2 rounds~12 and~14, did not exhibit the exponential leakage-current increase. As discussed in section~\ref{subsec:High_fluence_adaptations},
round~10 was an exception because limited dry-ice availability during the second irradiation period resulted in additional annealing.

\begin{figure}
	\captionsetup[subfigure]{aboveskip=-1pt,belowskip=-1pt}
	\centering
	\includegraphics[width=1\textwidth, trim=0 240 0 0, clip]{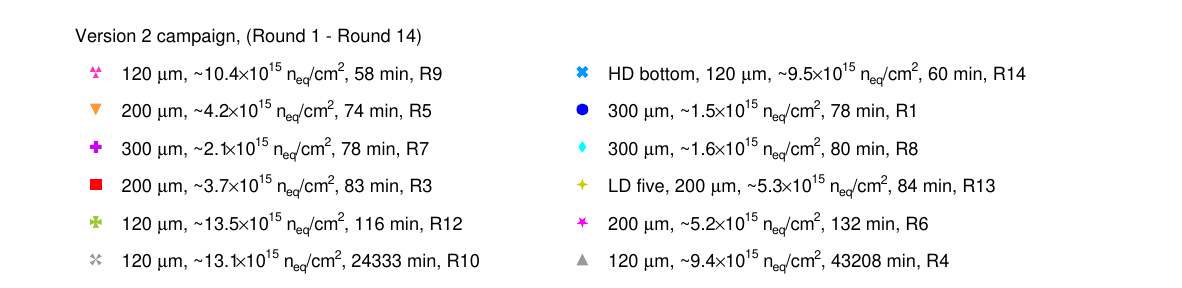}
	\begin{subfigure}[b]{0.49\textwidth}
		\includegraphics[width=1\textwidth, trim=8 0 21 0, clip]{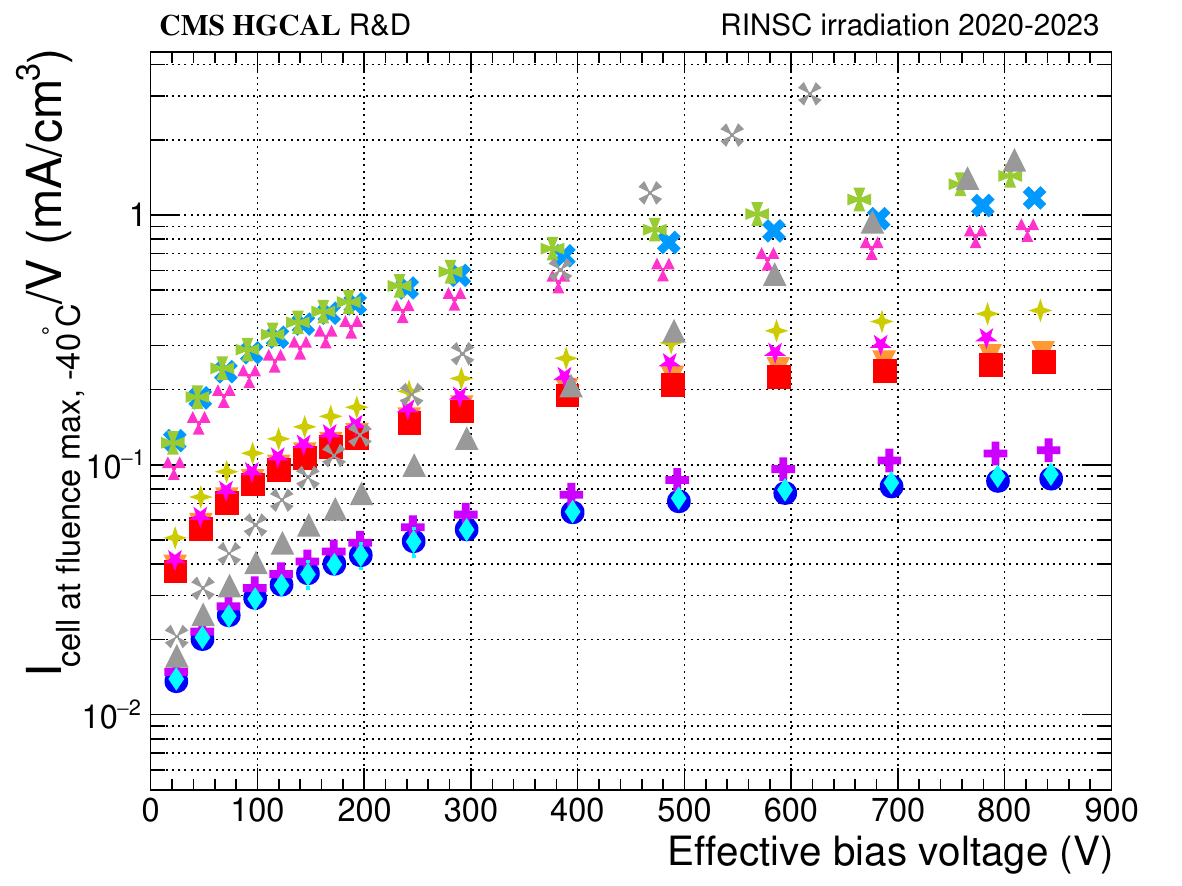}
		\subcaption{
		}
        \label{plot:IV_cell}
	\end{subfigure}
	\hspace{-0.2cm} 
    \begin{subfigure}[b]{0.49\textwidth}
		\includegraphics[width=1\textwidth, trim=8 0 21 0, clip]{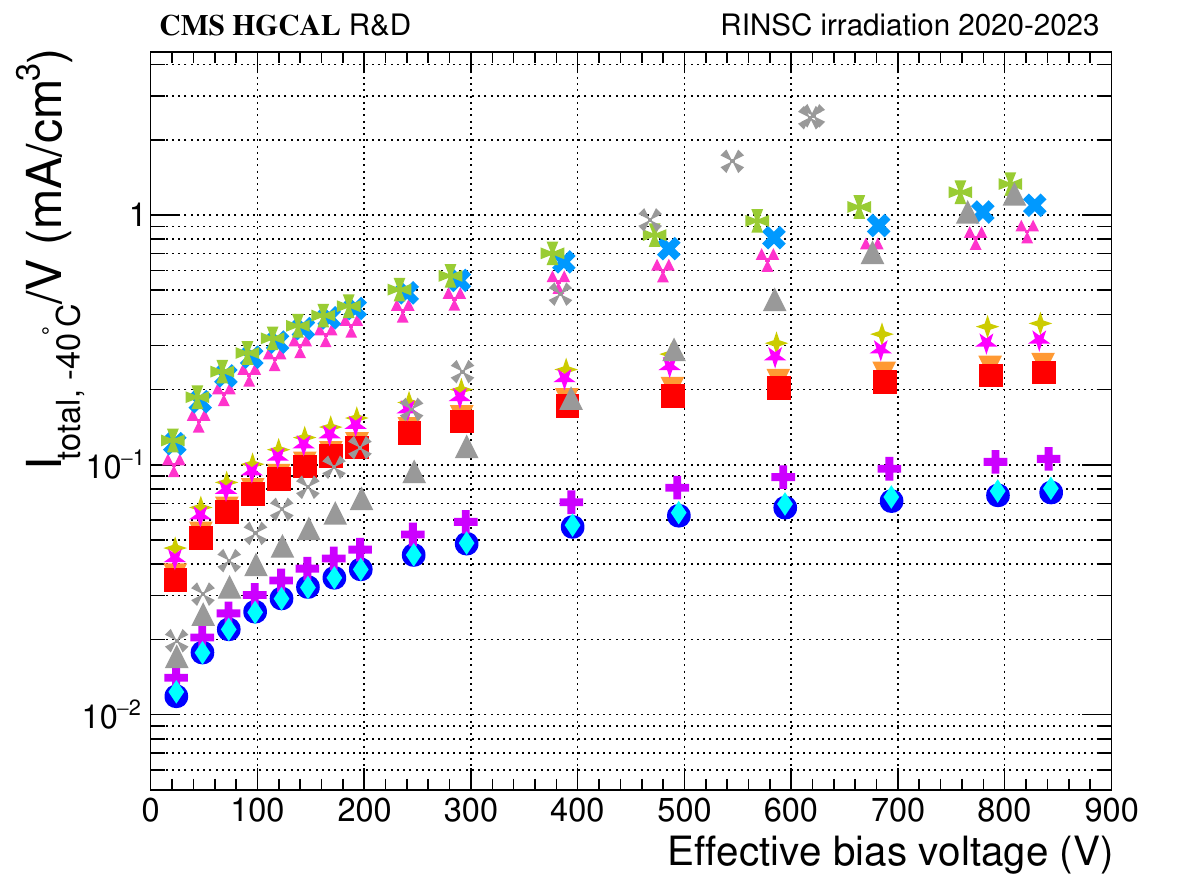}
		\subcaption{
		}		
        \label{plot:IV_total}
	\end{subfigure}

	\caption{
        Example IV curves for sensors from the version 2 irradiation campaign. a) Per-cell currents in current maximum. 
		b) Total current as seen by the power supply.
	}
\end{figure}

Figure~\ref{plot:IV_total} further shows that the total leakage current measured at the power supply exhibits the same exponential high-voltage behaviour
as the individual-cell leakage currents for sensors exposed to high fluence and prolonged annealing. This behaviour indicates that the effect is not limited to individual cells but applies to the entire sensor. 
When cumulative in-reactor annealing was reduced by splitting the irradiation into multiple steps, both the per-cell and
total-current IV characteristics recovered their stable diode-like behaviour.

These results indicate that limiting cumulative in-reactor annealing, for example by splitting high-fluence irradiation rounds,
is an effective strategy to suppress the exponential high-voltage increase in leakage current.
Minimizing excessive annealing and its resulting effects is crucial for
accurately assessing the full sensor performance in the detector, as the sensors will not experience
annealing times as prolonged as those observed during the irradiations at RINSC.

\section{Leakage current dependence on fluence}
\label{sec:Current_related_damage_rate}

\subsection{Cell current at fluence maximum}
\label{subsec:Cell_current_in_the_current_maximum}

The sensor leakage current contributes directly to the electronic noise and power dissipation of the detector~\cite{hgcal-tdr:2018}.
This section focuses on sensors irradiated to a maximum target fluence of $1.3 \times 10^{16}~\neqcm$, which is \SI{30}{\percent} above the expected End-of-Life fluence for the HL-LHC (see section~\ref{sec:intro}).
This provides a safety margin for evaluating sensor performance under extreme radiation conditions.

The leakage currents of the CE sensors were scaled to relevant temperatures for easier assessment using the temperature dependence:

\begin{equation}
\label{eq:factor}
I(T_{\text{ref}}) = I(T) \cdot \left(\frac{T_{\text{ref}}}{T}\right)^2 \cdot \exp\left(\frac{E_{\text{band}}}{2k_B} \left(\frac{1}{T} - \frac{1}{T_{\text{ref}}} \right) \right),
\end{equation}
where $E_{\text{band}} = 1.21$~eV is the effective bandgap energy~\cite{Dawson:2021}, and $k_B$ is the Boltzmann constant.
This relation models the temperature dependence of the leakage current in silicon and is widely used in high-energy physics~\cite{moll:SiDamages}.
Its validity for CE sensors over the relevant temperature range has been confirmed in~\cite{Acar_2023} for fluences around $1 \times 10^{15}~\neqcm$.
The accuracy of this scaling method at higher fluences, such as $1 \times 10^{16}~\neqcm$, is further discussed in appendix~\ref{appendix:scaling}.

Table~\ref{table:limits} summarizes the leakage current estimates at both \SI{-35}{\celsius} and \SI{-30}{\celsius}, for cells at current maximum and the total current of the sensor as seen by the power supply.
The target temperature of \SI{-35}{\celsius} corresponds to the planned CE operating condition.
However, measurements suggest that certain regions of the sensors may reach temperatures as high as \SI{-30}{\celsius}, due to thermal resistance in the cooling setup as well as heat generated by components on the front-end PCB layer. 
Therefore, the scenario of sensor cells operating at \SI{-30}{\celsius} was also evaluated.

While the per-cell leakage current remains within specification under all conditions, the total sensor current exceeds the design limit at \SI{-30}{\celsius}, emphasizing the need for effective detector cooling.
It is also worth noting that partial sensors produce lower total leakage currents due to their reduced number of cells.

\begin{table}[h]
	\centering
	\caption{Estimated per-cell and total leakage current for different CE detector operating scenarios, for a sensor irradiated to a target fluence of $1.3 \times 10^{16}~\text{n}_{\text{eq}}/\text{cm}^2$ (version 2 Round 12).}
	\label{table:limits}
	\begin{tabular}{lcccc}
		\toprule
		\textbf{Temperature} & 
		\multicolumn{2}{c}{\textbf{\boldmath$V_{\text{bias}} = \SI{600}{\volt}$}} & 
		\multicolumn{2}{c}{\textbf{\boldmath$V_{\text{bias}} = \SI{800}{\volt}$}} \\
		& $I_{\text{cell at fluence max}}~[\si{\micro\ampere}]$ & $I_{\text{total}}~[\si{\milli\ampere}]$
		& $I_{\text{cell at fluence max}}~[\si{\micro\ampere}]$ & $I_{\text{total}}~[\si{\milli\ampere}]$ \\
		\midrule
		\SI{-35}{\celsius} & 13.7 & 5.45 & 18.45 & 7.26 \\
		\SI{-30}{\celsius} & 26.21 & 10.43 & 35.27 & 13.88 \\
		\midrule
		\textbf{Spec. Limit (Per Cell)} & \textbf{50} & – & \textbf{50} & – \\
		\textbf{Spec. Limit (Total)} & – & \textbf{10} & – & \textbf{10} \\
		\bottomrule
	\end{tabular}
\end{table}

\subsection{Current-related damage factor}
\label{subsec:Current_damage coefficient}

The current-related damage factor ($\alpha$) quantifies the proportionality between the radiation-induced increase in leakage current ($\Delta I$), 
normalized to the sensor volume ($V$), and the particle fluence ($\Phi$). The~$\alpha$ parameter is independent of the type, resistivity, and impurity content of the silicon material used~\cite{moll:SiDamages,MOLL199987}. 
It is defined as:

\begin{equation} \label{eq:alpha} \alpha = \frac{\Delta I}{V \cdot \Phi}, \end{equation}

Table~\ref{table:alphas} presents the $\alpha$ values obtained in this study scaled to different temperatures for comparison with the reference values from the previous studies on silicon sensors, which were also included in the table.
The $\alpha$ factor was extracted from the leakage current data through a fit. The leakage current for this analysis was determined by averaging the three highest-leakage current cells within each sensor. 
The data presented corresponded to samples annealed for a total equivalent of $\SI[separate-uncertainty = true]{80(20)}{\minute}$ at \SI{60}{\celsius}, unless stated otherwise.

To extend the dataset analysed in this study, results from version 1 CE sensors campaign were included alongside those from the version 2 campaign.
However, the $\alpha$ value obtained in this study is not directly comparable to the value reported for version 1 sensors in ref.~\cite{Acar_2023}, as a different fluence estimation method has since been used (see section~\ref{subsec:Fluence_assessment}). 
Therefore, the $\alpha$ value of the version 1 campaign was recalculated in this study using the fluence derived from the irradiation time and is presented in table~\ref{table:alphas}.

\begin{figure}
	\captionsetup[subfigure]{aboveskip=-1pt,belowskip=-1pt}
	\centering
    
    \begin{subfigure}[c]{0.49\textwidth}
		\includegraphics[width=0.999\textwidth, trim=17 0 18 0, clip]{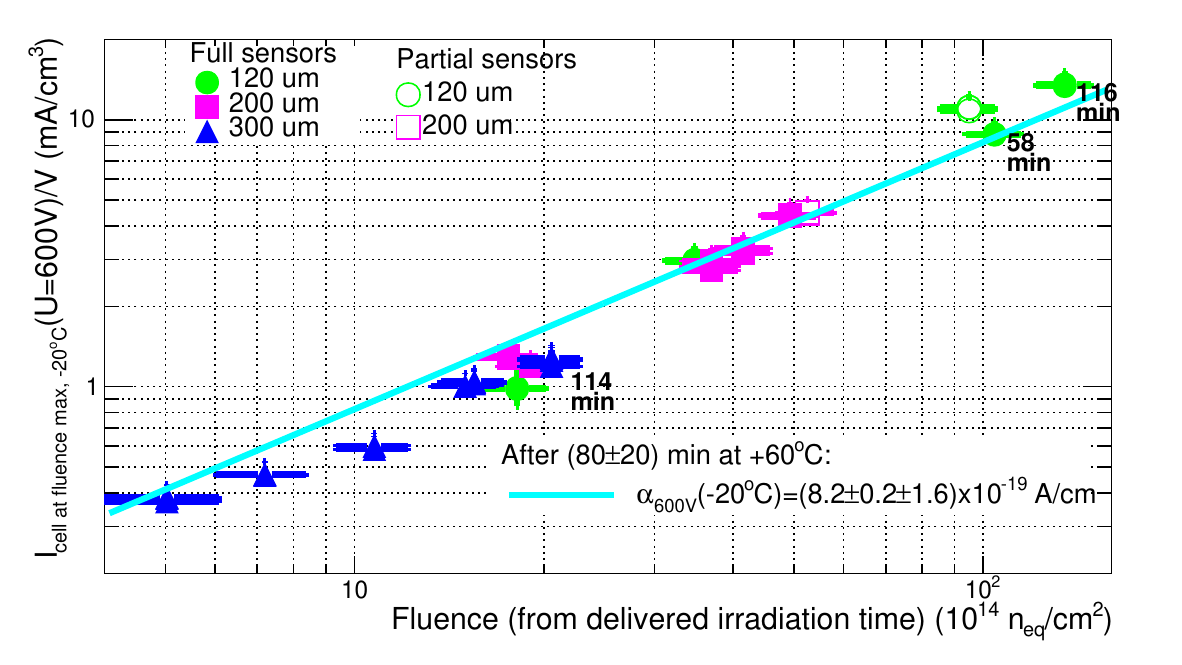}
		\subcaption{
		}		
        \label{plot:alpha_versions_together}
	\end{subfigure}
	\begin{subfigure}[c]{0.49\textwidth}
		\includegraphics[width=0.999\textwidth, trim=17 0 18 0, clip]{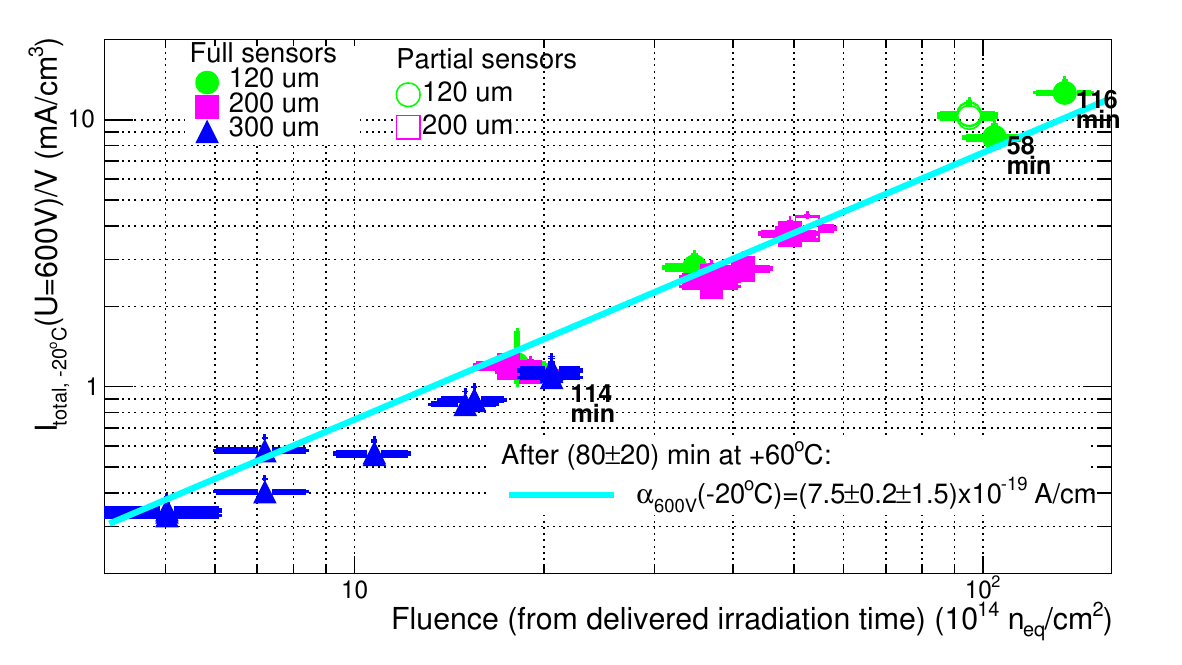}
		\subcaption{
		}		
        \label{plot:Alpha_total}
	\end{subfigure}
	\caption{
		a) Cell leakage current at current maximum versus delivered fluence, based on combined data from the version 1 and version 2 campaigns. b) Total leakage current as seen by the power supply versus delivered fluence.
		}
\end{figure}

The version 2 campaign shows an $\alpha$ value approximately 16$~\%$ higher than that of the version 1 campaign at \SI{-20}{\celsius}.
The available measurements do not allow the origin of this discrepancy to be determined conclusively. 
A possible contributing factor may be an incomplete estimation of the annealing time in parts of the version 
1 campaign, however, this cannot be verified from the available data. 
Therefore, this difference was conservatively incorporated into the systematic uncertainty. The total systematic uncertainty on $\alpha$ was conservatively estimated to be 20$~\%$.
After this adjustment, the results from the two campaigns are consistent within the total uncertainties, and the datasets were combined for the final fit shown in figure~\ref{plot:alpha_versions_together}.

In table~\ref{table:alphas}, the first quoted uncertainty is statistical, and the second is systematic. The statistical uncertainty includes contributions from the fit, fluence estimation, annealing-time estimation, 
temperature uncertainty during the IV measurements, and sensor thickness variation. The systematic uncertainty accounts for the observed discrepancy between the version 1 and version 2 campaigns, as well as facility-dependent effects related to irradiation conditions and fluence determination. 
The total uncertainties are obtained by adding the individual contributions in quadrature within each category.

\begin{table}[h]
	\centering
	\small
	\caption{Comparison of the $\alpha$ values obtained in this study, scaled to different temperatures using the normalization relation presented in eq.~\ref{eq:factor}, 
	with reference $\alpha$ values reported in other studies. For the CE sensor results, the first quoted uncertainty is statistical, and the second is systematic (see text for details).}
	\label{table:alphas}	
	\begin{threeparttable}
	\begin{tabular}{lccc}
		\textbf{Data} 
		& $\alpha_{\SI{-35}{\celsius}}~\times 10^{-19}~\ensuremath{\mathrm{A}/\mathrm{cm}}\xspace$ 
		& $\alpha_{\SI{-20}{\celsius}}~\times 10^{-19}~\ensuremath{\mathrm{A}/\mathrm{cm}}\xspace$ 
		& $\alpha_{\SI{20}{\celsius}}~\times 10^{-17}~\ensuremath{\mathrm{A}/\mathrm{cm}}\xspace$ \\
		\hline
		CE sensors, version 1\tnote{A,B} & $1.1\pm0.0\pm0.2$ & $7.4\pm0.3\pm1.5$ & $4.4\pm0.2\pm0.9$\\
		CE sensors, version 2\tnote{A,B} & $1.3\pm0.0\pm0.3$ & $8.6\pm0.2\pm1.7$ & $5.1\pm0.1\pm1.0$\\
		CE sensors, combined\tnote{A,B,C} & $1.3\pm0.0\pm0.3$  & $8.2\pm0.2\pm1.6$ & $4.7\pm0.1\pm1.0$\\
		CE sensors, total current\tnote{A,B,C} & $1.2\pm0.0\pm0.2$ & $7.5\pm0.2\pm1.5$ & $4.4\pm0.1\pm0.9$\\
		CE diodes (JSI)~\cite{Kieseler_2023}\tnote{D,E} &  & $6.8\pm0.1\pm0.7$ & \\
		p-type sensors~\cite{moll:SiDamages,MOLL199987}\tnote{E,F} &  &  & $3.99\pm0.03$ \\
		\hline
	\end{tabular}
    \begin{tablenotes}\footnotesize
	\item[A] Irradiated at RINSC.
	\item[B] Measured at the bias voltage \SI{600}{\volt} and at fluence maximum.
	\item[C] “Combined” refers to a fit using CE sensor data from both version 1 and version 2 campaigns.
	\item[D] Irradiated at JSI in Ljubljana.
	\item[E] Measured at the depletion voltage.
	\item[F] Irradiated mostly at Physikalisch-Technische Bundesanstalt (PTB) in Braunschweig.
    \end{tablenotes}
    \end{threeparttable}
\end{table}

The combined CE sensor dataset irradiated at RINSC exhibits $\alpha$ values that are systematically higher than those of CE diodes (21$~\%$) and other p-type sensors (18$~\%$) irradiated at other facilities, as shown in table~\ref{table:alphas}. 
Variations in neutron flux and exact fluence estimation between different facilities cannot be neglected.
Since JSI is a reference facility, the difference in $\alpha$ value between the CE sensors irradiated at RINSC and the CE diodes irradiated at JSI was included in the estimation of the systematic uncertainty of the result.
This uncertainty accounts for irradiation-related effects at RINSC and the limited understanding of the annealing process.
Also, the reference values were obtained at the 
depletion voltage (measured at the frequency of \SI{10}{\kHz}), while the bias voltage of \SI{600}{\volt} was utilized in this study. 
The depletion voltage is lower than
the mentioned bias voltage even for the samples of \SI{300}{\micro\meter} thickness~\cite{Kieseler_2023}.

The volume-normalized total current versus fluence is compatible with the cell current taken at the fluence maximum, as shown in figure~\ref{plot:Alpha_total}. The slightly reduced $\alpha$ value is related to the on average lower 
fluence exposure of the full sensor with respect to the cells in fluence maximum. If the average current of the full cells is used instead of the maximum, an $\alpha$ value of 
$7.5\pm0.2\pm1.5~\times 10^{-19}~\ensuremath{\mathrm{A}/\mathrm{cm}}$ is obtained, which is closer to the $\alpha$ value for the total current.

\subsection{Validation of fluence assessment procedure}
\label{subsec:Validation_of_fluence_assessment_procedure}

For the $\alpha$ factor analysis, the use of an accurate fluence value is crucial. This study compares three fluence estimation methods: 
(i) fluence estimation from dosimetry sensors, (ii) target fluence, and (iii) fluence estimation from irradiation time. 
These methods were introduced in section~\ref{subsec:Fluence_assessment}.

In the version 1 campaign analysis reported in ref.~\cite{Acar_2023}, fluence was estimated using dosimetry objects. However, the number and type of dosimetry sensors varied between campaigns (and even between rounds within a campaign)
making this method unsuitable for reliable comparison. As discussed in section~\ref{subsec:Fluence_assessment}, the target fluence is derived from a time conversion factor, which was modified between campaigns. For example, the version 2 campaign involved longer
irradiation times for the same target fluence as version 1, leading to inaccurate target fluence estimates for some version 1 rounds. This inconsistency excludes target fluence as a reliable option for fluence assessment. 
Since the reactor was operated at stable power during irradiation (see section~\ref{subsec:Fluence_assessment}), the fluence estimation based on delivered irradiation time, calculated from the reactor power curves, is both reliable and reproducible across campaigns.

\begin{figure}
	\captionsetup[subfigure]{aboveskip=-1pt,belowskip=-1pt}
	\centering
	\includegraphics[width=0.999\textwidth]{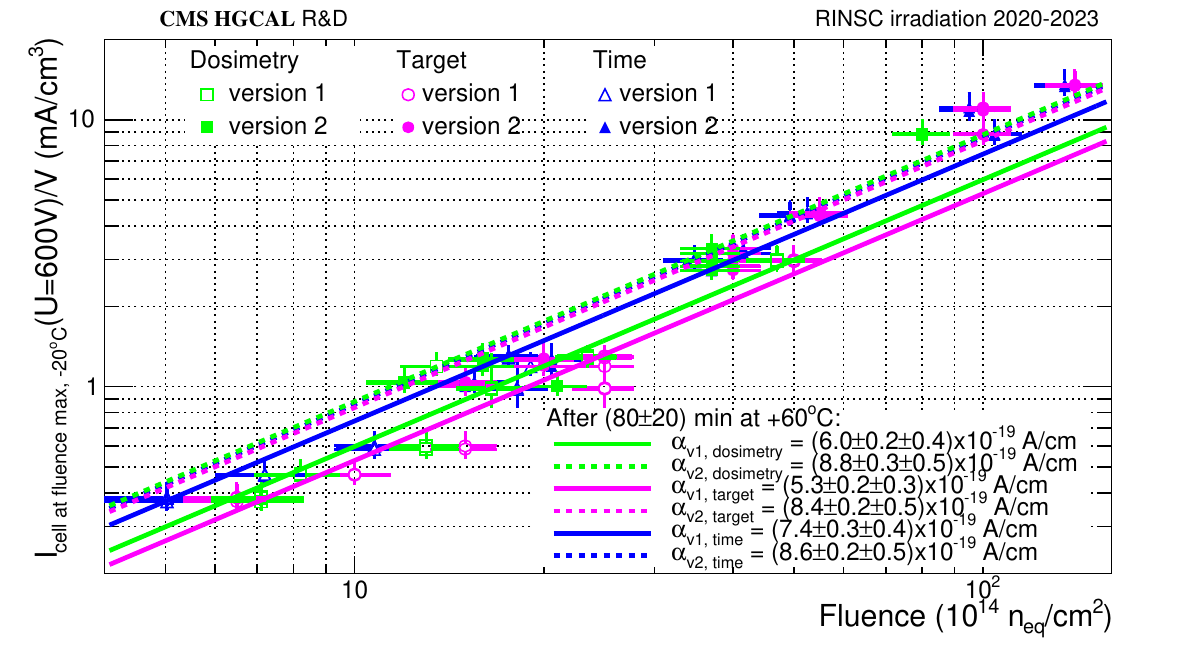}
	\caption{
		Comparison of per-cell leakage current at sensor current maximum using three fluence evaluation methods.
	}
	\label{plot:alpha_600V_thickness_delivered_irradiationTimeV1V2separately_Fluence_Estimation}
\end{figure}

These considerations are further supported by a comparison of the fitted leakage currents and extracted $\alpha$ values for each estimation method.  
As shown in figure~\ref{plot:alpha_600V_thickness_delivered_irradiationTimeV1V2separately_Fluence_Estimation}, a systematic offset between the version 1 and version 2 campaigns is observed when using dosimetry-based fluence estimates.  
An even larger offset is seen when using the target fluence method. In contrast, when fluence is estimated from delivered irradiation time, the fit lines from both campaigns align more closely, and the extracted $\alpha$ values agree within their respective uncertainties.  
Therefore, the analyses in this work use fluence values based on delivered irradiation time, applying a consistent conversion factor to both campaigns.

Additionally, it is worth mentioning that figure~\ref{plot:alpha_600V_thickness_delivered_irradiationTimeV1V2separately_Fluence_Estimation} also illustrates that the current-related damage rates $\alpha_{\SI{-20}{\celsius}}$ 
for version 1 sensors range from \mbox{$5.3 \times 10^{-19}~\ensuremath{\mathrm{A}/\mathrm{cm}}$} to \mbox{$7.4 \times 10^{-19}~\ensuremath{\mathrm{A}/\mathrm{cm}}$}, 
while for version 2 sensors, the range is narrower, from \mbox{$8.4 \times 10^{-19}~\ensuremath{\mathrm{A}/\mathrm{cm}}$} to \mbox{$8.6 \times 10^{-19}~\ensuremath{\mathrm{A}/\mathrm{cm}}$}. 
This suggests that fluence assessment methods have improved in the version 2 campaign, but still retain some degree of uncertainty.
These improvements can be attributed to better sensor placement, an increased number of dosimetry objects, and an optimized time conversion factor (see section~\ref{subsec:Fluence_assessment}).

\section{Leakage current versus temperature}
\label{sec:Temperature}

\subsection{Extraction of activation energy}
\label{subsec:Activation_energy}

\begin{figure}
	\captionsetup[subfigure]{aboveskip=-1pt,belowskip=-1pt}
	\centering
	\begin{subfigure}[b]{0.49\textwidth}
		\includegraphics[width=0.999\textwidth]{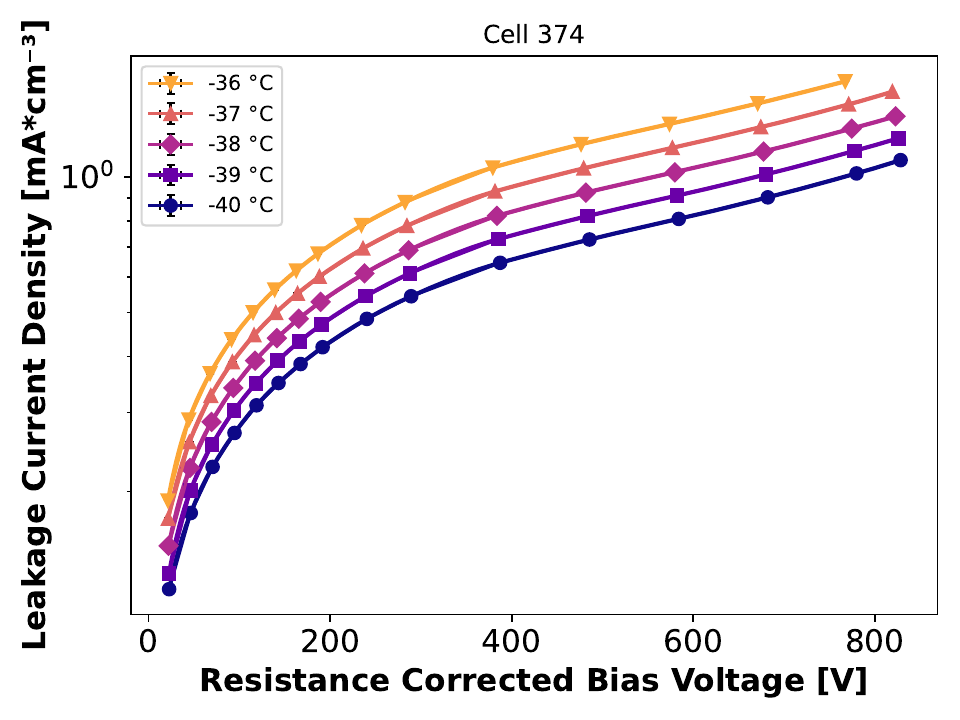}
		\subcaption{
		}
        \label{plot:temperature_leakage}
	\end{subfigure}
    \hfill
    \begin{subfigure}[b]{0.49\textwidth}
		\includegraphics[width=0.999\textwidth]{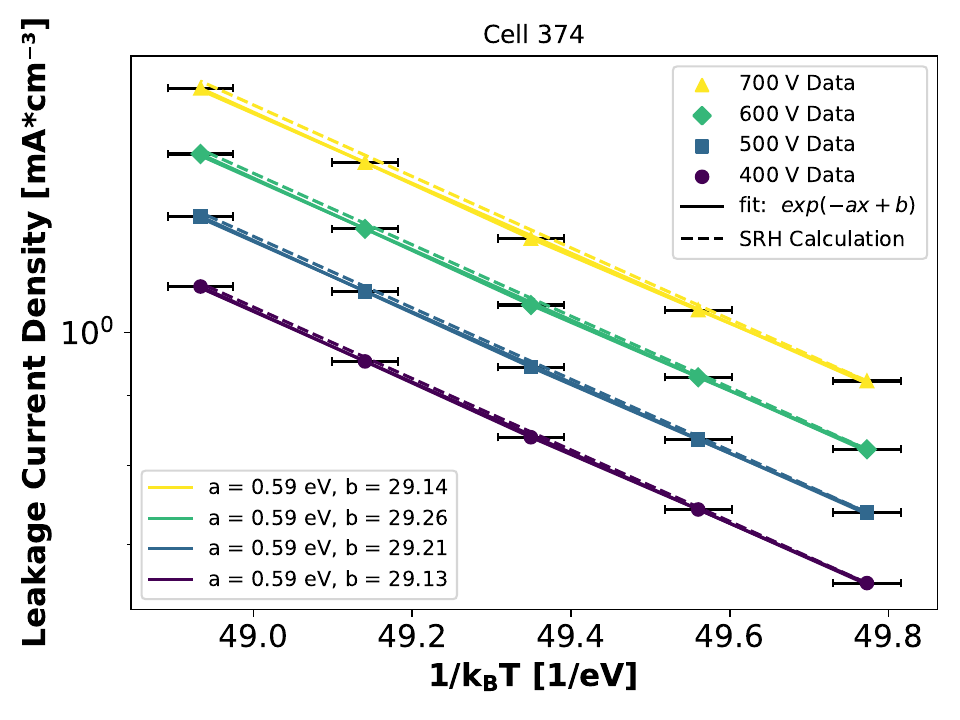}
		\subcaption{
		}		
        \label{plot:arrhenius}
	\end{subfigure}

	\caption{
        a) Per cell (number 374) leakage current scaled with cell volume. 
        Cell 374 is a full cell of a partial sensor of high density granularity 
        and \SI{120}{\micro\meter} thickness, irradiated to a delivered fluence of 
        $9.5 \times 10^{15}~\neqcm$ (version 2 Round 14). b) Arrhenius curves at different interpolated bias voltages for 
        the same cell of the same sensor as in a).
		}
\end{figure}

The temperature dependence of the leakage current in neutron-irradiated CE sensors was investigated in a narrow interval between \SI{-40}{\celsius} and \SI{-36}{\celsius}. This range was selected based on 
practical limitations of the measurement setup. The lower limit of \SI{-40}{\celsius} corresponds to the lowest temperature reachable by the cold chuck~\cite{att-chucks}, 
while measurements above \SI{-36}{\celsius} posed a risk of exceeding the safe operational current limits of the measurement setup, and some measurement points were not recorded specifically at the bias
voltage of \SI{800}{\volt}, as seen in figure~\ref{plot:temperature_leakage}. 
Within the accessible range, the temperature was incremented in \SI{1}{\celsius} steps to ensure precise extraction of the activation energy.

The activation energy ($E_A$) was extracted by analysing the temperature dependence of the leakage current density,
following an Arrhenius-type behaviour, as illustrated in figure~\ref{plot:arrhenius}. For sensors irradiated to various fluences, activation energies in the range 0.58–0.63 eV were measured.
These values are consistent with previous findings reported in ref.~\cite{Obreja}.

\subsection{Comparison with Shockley Read Hall theory}
\label{subsec:Shockley_Read_Hall_theory}

The Shockley Read Hall (SRH) theory describes the temperature-leakage current dependence $I(T)$ for bulk leakage current as follows:
\begin{equation}
    \label{eq:Arrhenius}
    I(T) \approx T^{2} \exp\left(-\frac{E_{A}}{2k_B T}\right).
\end{equation}
This theoretical model, derived from the migration rate of defects in a material~\cite{Moll:300958}, 
helps distinguish between leakage current contributions from bulk recombination centres and surface effects, such as charge trapping in the silicon oxide layer~\cite{AChilingarov_2013}.

As shown in figure~\ref{plot:temperature_leakage}, the leakage current increases with temperature, as expected from the Shockley-Read-Hall model~\cite{AChilingarov_2013,Spieler:2005si}.
The activation energy values extracted in section~\ref{subsec:Activation_energy} correspond to an equivalent band gap energy range of 1.16–1.26 eV, using the relation:

\begin{equation} E_{band} = 2E_{A}, \end{equation}
as described in ref.~\cite{Grove:1967}. This range aligns closely with the widely accepted band gap energy value of $E_{band}=1.21$ eV,
as also noted in appendix~\ref{appendix:scaling}.

The agreement between the measured activation energies and the SRH prediction indicates that bulk recombination centres, rather than surface defects, are the dominant source of leakage current in neutron-irradiated CE sensors.
The absence of significant deviations from the SRH model further suggests that contributions from surface-related leakage mechanisms (e.g., silicon oxide charge trapping) are negligible.
\section{Summary and outlook}

\label{sec:Summary}

This study investigates the behaviour of CE silicon sensors exposed to fluences in range between $5 \times 10^{14}~\neqcm$ and $1.3 \times 10^{16}~\neqcm$. The results show that the leakage current profiles are smooth for all sensors, 
with no instabilities related to sensor layouts or proximity to internal HV structures in partial sensors.

Most sensors exhibit a typical diode-like IV dependence. However, sensors exposed to prolonged annealing and high fluences developed a
pronounced exponential increase in leakage current at high voltages. This behaviour was mitigated by splitting high-fluence irradiation rounds to
limit cumulative in-reactor annealing, demonstrating that controlling the annealing history suppresses the exponential increase in
leakage current and preserves stable diode-like IV characteristics.

The measured leakage current at a fluence approximately 30$~\%$ above the expected HL-LHC End-of-Life fluence confirms that CE sensors perform as expected under operational conditions.
At the planned operating temperature of \SI{-35}{\celsius}, both the per-cell and total sensor leakage currents remain within the limits set by the detector design and the readout chip.
However, when assuming a higher sensor temperature of \SI{-30}{\celsius}, the total leakage current would exceed the specification limit, showing the importance of efficient cooling to ensure full compliance at extreme fluences.

The leakage current increases with fluence, following the expected trend. We observed that partial sensors with internal 
HV lines exhibit leakage current behaviour consistent with that of full sensors. 
Additionally, HD sensors irradiated up to $1.3 \times 10^{16}~\neqcm$ follow the anticipated leakage current trend.

The current-related damage factor values obtained in the previous and current irradiation campaigns are consistent within the total uncertainties, justifying their combination into a single dataset.
A comparison of single-cell and total sensor currents from both full and partial sensors reveals consistent behaviour across all sensor types. 
 However, the consolidated CE sensor $\alpha$ values are systematically higher than those reported in the literature, likely due to variations in neutron flux, fluence estimation across facilities, and bias voltage differences.

Finally, this work presents the first measurement of the temperature dependence of the IV characteristics of CE silicon 
sensors after high-fluence neutron irradiation. The measured activation energies are consistent with theoretical expectations, 
indicating that bulk recombination centres, rather than surface defects, dominate the leakage current in irradiated CE silicon sensors.

Neutron irradiation at RINSC has proven essential for assessing the bulk radiation hardness of large-area silicon sensors for the CE detector.
For future neutron irradiations at RINSC, we recommend deploying additional temperature sensors to improve the monitoring of sensor annealing and to minimize uncertainties in the assessment of 
in-reactor annealing. Specifically, we suggest placing at least two RTDs on the front and two on the back of the puck, ideally positioned directly adjacent to the silicon sensors.
Further improvements could be achieved by directly measuring the horizontal temperature distribution across the sensor area, 
for which an array of temperature sensors distributed across the wafer is proposed.

Additionally, ensuring consistent sensor orientation (rotation) relative to the reactor during irradiations is recommended. 
  This would facilitate a more accurate assessment of annealing and fluence profiles, particularly for high-fluence rounds (>$1.0 \times 10^{16}~\neqcm$) that are split into multiple irradiation parts.
  
\acknowledgments

We thank the staff at the Rhode Island Nuclear Science Center for their support during the preparation 
and execution of the neutron irradiation of the CE silicon pad sensor prototypes. 
The EP-DT and the former EP-LCD group at CERN have developed essential infrastructure, 
such as the ARRAY system including the associated data acquisition software, and have co-financed 
the acquisition of the utilised cold-chuck probe station at CERN. Without their input, this R\&D 
milestone towards the realisation of this novel calorimeter would not have been possible.
We thank Zachary Zawisza for his analysis of the leakage-current temperature dependence.
We are thankful for the technical and administrative support at CERN and at other CMS institutes and 
thank the staff for their contributions to the success of the CMS effort. We acknowledge the enduring support 
provided by the following funding agencies and laboratories: BMBWF and FWF (Austria); 
CERN; CAS, MoST, and NSFC (China); MSES and CSF (Croatia); CEA, CNRS/IN2P3 and P2IO LabEx (
    ANR10-LABX-0038) (France); SRNSF (Georgia); BMBF, DFG, and HGF (Germany); GSRT (Greece); 
    DAE and DST (India); MES (Latvia); MOE and UM (Malaysia); MOS (Montenegro); PAEC (Pakistan); 
    FCT (Portugal); JINR (Dubna); MON, RosAtom, RAS, RFBR, and NRC KI (Russia); MoST (Taipei); 
    ThEP Center, IPST, STAR, and NSTDA (Thailand); TUBITAK and TENMAK (Turkey); STFC (United Kingdom); and DOE (USA).

\appendix
\section*{Appendix}
\addcontentsline{toc}{section}{Appendix} 

\section{Pre-cooling of the aluminium cylinder}
\label{appendix:pre-cooling}

Before the second half of round 14 in the version 2 campaign, the aluminium cylinder was kept in a freezer due to radiation safety concerns, as it still 
showed residual radioactivity from the first half of the irradiation. 
This raised a discussion about whether pre-cooling could reduce in-reactor annealing. According to table~\ref{tab:summary}, pre-cooling reduced the equivalent time at \SI{60}{\celsius} by 31$~\%$ in the 
front measurement. However, a comparable reduction of 34$~\%$ was observed in round 12 without pre-cooling. Given the significant effort required for pre-cooling, 
including an extra trip to the reactor, and the lack of substantial impact on in-reactor annealing, this method is not recommended for future campaigns.

\section{Leakage current temperature scaling}
\label{appendix:scaling}

Since the leakage current $I(T_{\text{ref}})$ in silicon sensors is strongly temperature-dependent, it must be scaled to a common reference temperature ($T_{\text{ref}}$) to ensure consistent comparison across different measurement conditions.
This correction is applied using an exponential scaling law derived from the temperature dependence of the leakage current in silicon~\cite{AChilingarov_2013}:

\begin{equation}
	\label{eq:scaling}
	I(T_{\text{ref}}) = I(T) \cdot c(T, T_{\text{ref}}),
\end{equation}
where the scaling factor $c(T, T_{\text{ref}})$ is given by:

\begin{equation}
	\label{eq:factor1}
	c(T, T_{\text{ref}}) = \left(\frac{T_{\text{ref}}}{T}\right)^2 \cdot \exp\left( \frac{E_{\text{band}}}{2k_B} \left( \frac{1}{T} - \frac{1}{T_{\text{ref}}} \right) \right).
\end{equation}

Here, $E_{\text{band}} = 1.21$~eV is the effective silicon bandgap energy~\cite{Dawson:2021}, and $k_B$ is the Boltzmann constant.
This expression is widely adopted in the high-energy physics community for leakage current normalization~\cite{moll:SiDamages}.

The validity of this scaling law for CE silicon sensors at moderate fluences has been confirmed in~\cite{Acar_2023}. To evaluate its accuracy at high fluences (e.g., $1 \times 10^{16}~\neqcm$), 
scaled leakage current values were compared to direct measurements taken over a narrow temperature range between \SI{-40}{\celsius} and \SI{-36}{\celsius}.

The comparison in table~\ref{table:sensor_temperature_scaling} confirms that the leakage current increases with temperature, as expected~\cite{Spieler:2005si}, and that the scaling law remains accurate within this temperature interval.
At \SI{600}{\volt} and \SI{-36}{\celsius}, the average absolute deviation between measured and scaled values was approximately \SI{0.50}{\micro\ampere} at the cell level and \SI{0.11}{\milli\ampere} for the total current.

\begin{table}[h]
    \centering
    \small
    \caption{Per-cell leakage current measurements and scaled results for CE sensors across temperatures at two voltage settings.}
    \label{table:sensor_temperature_scaling}
    \begin{tabular}{c c c c c c c c}
        \toprule
        \textbf{Sensor} & \textbf{Voltage} & \textbf{Type} 
        & \textbf{\SI{-40}{\celsius}} & \textbf{\SI{-39}{\celsius}} & \textbf{\SI{-38}{\celsius}} 
        & \textbf{\SI{-37}{\celsius}} & \textbf{\SI{-36}{\celsius}}\\
        \midrule

        \multirow{6}{*}{\textbf{V2, R14}}& \multirow{3}{*}{600 V}
            & $I_{\text{cell, measured}}~[\si{\micro\ampere}]$  & 5.83 & 6.55 & 7.48 & 8.38 & 9.53\\
            & & $I_{\text{cell, scaled}}~[\si{\micro\ampere}]$ & - & 6.69 & 7.67 & 8.77 & 10.03\\
            & & $I_{\text{cell, difference}}~[\si{\micro\ampere}]$ & - & -0.14 & -0.19 & -0.39 & -0.50 \\

        \cline{2-8}

        & \multirow{3}{*}{800 V}
            & $I_{\text{cell, measured}}~[\si{\micro\ampere}]$ & 7.40 & 8.32 & 9.49 & - & -\\
            & & $I_{\text{cell, scaled}}~[\si{\micro\ampere}]$ & - & 8.49 & 9.72 & 11.13 & 12.72\\
            & & $I_{\text{cell, difference}}~[\si{\micro\ampere}]$ & - & -0.17 & -0.23 & - & - \\

		\midrule
		\multirow{6}{*}{\textbf{V2, R14}}& \multirow{3}{*}{600 V}
            & $I_{\text{total, measured}}~[\si{\milli\ampere}]$ & 1.37 & 1.55 & 1.76 & 1.97 & 2.25\\
            & & $I_{\text{total, scaled}}~[\si{\milli\ampere}]$ & - & 1.57 & 1.80 & 2.06 & 2.36\\
            &                 & $I_{\text{total, difference}}~[\si{\milli\ampere}]$ & - & -0.02 & -0.04 & -0.09 & -0.11\\

        \cline{2-8}

        & \multirow{3}{*}{800 V}
            & $I_{\text{total, measured}}~[\si{\milli\ampere}]$ & 1.74 & 1.96 & 2.24 & 2.50 & 2.85\\
            & & $I_{\text{total, scaled}}~[\si{\milli\ampere}]$ & - & 2.00 & 2.29 & 2.62 & 3.00\\
            &                 & $I_{\text{total, difference}}~[\si{\milli\ampere}]$ & - & -0.04 & -0.05 & -0.12 & -0.15\\

        \bottomrule
    \end{tabular}
\end{table}

\bibliographystyle{JHEP}
\bibliography{Bib/Bib}

\end{document}